\documentclass[12pt,reqno]{article}

\usepackage[T1]{fontenc}
\usepackage[utf8]{inputenc}
\usepackage{newpxtext,newpxmath} 
\usepackage{microtype}
\usepackage[fontsize=12pt]{fontsize}

\usepackage{setspace}

\usepackage[right=1.08in,left=1.08in,top=1.08in,bottom=1.08in]{geometry}

\usepackage{amsmath,amsthm,graphicx}
\usepackage{subcaption}
\usepackage{booktabs,multirow,threeparttable,array,adjustbox,longtable}
\usepackage{makecell}
\usepackage[dvipsnames]{xcolor}
\usepackage{float}

\usepackage{natbib}
\usepackage{xurl}
\usepackage{hyperref}
\hypersetup{
    colorlinks=true,
    linkcolor=red!75!black,
    citecolor=blue!75!black,
    filecolor=magenta,      
    urlcolor=red!75!black,
}
\usepackage[noabbrev]{cleveref}

\usepackage[linesnumbered,ruled,vlined]{algorithm2e}

\usepackage{sectsty}
\usepackage{bm}
\usepackage{fancyhdr}

\newtheorem{proposition}{Proposition}
\newtheorem{remark}{Remark}

\title{
\vspace*{-2.5cm} \hspace*{-0.5cm}
\textsc{Temporal Disaggregation of GDP: \\ When Does Machine Learning Help?}
\thanks{A previous version of this paper was circulated under the title ``Machine Learning-Based Estimation of Monthly GDP.'' I am grateful to Gunnar Heins, Sooji Kim, and Yajie Xu for their valuable comments, and to the participants of the 2025 Workshop in Applied and Theoretical Economics (WATE) for helpful suggestions. All errors are my own. Replication package and data: \href{https://github.com/Yonggeun-Jung/monthly_gdp}{https://github.com/Yonggeun-Jung/monthly\_gdp}. First version: June, 2025.}
}

\author{
\textsc{Yonggeun Jung}\protect\thanks{Department of Economics, University of Florida. Email: \href{mailto:yonggeun.jung@ufl.edu}{yonggeun.jung@ufl.edu}.}
}

\date{ \vspace*{0.5cm} This version: April, 2026} 

\begin{document}
\bgroup

\begin{singlespace}
\maketitle

\begin{abstract}
\noindent
We propose a modular framework for temporal disaggregation of quarterly GDP into monthly frequency, in which the regression step accommodates any supervised learning model while Mariano-Murasawa reconciliation enforces quarterly consistency. Comparing Chow-Lin, Elastic Net, XGBoost, and a Multi-Layer Perceptron across four countries, we find that regularization, not nonlinearity, drives the gains: Elastic Net achieves $R^2 = 0.87$ for the United States when lagged indicators are included, while nonlinear models cannot overcome the variance cost of small quarterly samples. We formalize this tradeoff through regime-switching bias and ridge-regularization results.
\end{abstract}

\end{singlespace}
\vspace{0.25em}
\thispagestyle{empty} 

\clearpage
\egroup
\setcounter{page}{1}

\onehalfspacing

\section{Introduction \label{sec:intro}}

Real gross domestic product (GDP) is a core economic indicator, widely valued for its intuitive and comprehensive representation of economic activity within a given region. While GDP is traditionally published at a quarterly frequency, the increasing demand for timely and granular economic insights has led to growing interest in both nowcasting and high-frequency estimation \citep{mitchell2005indicator, giannone2008nowcasting, jansen2016forecasting, chu2023comparing, ghosh2023machine, richardson2021nowcasting}.

In particular, the expanding availability of high-frequency data across various domains has intensified efforts to estimate monthly GDP \citep{mitchell2005indicator, koop2023reconciled, koop2021nowcasting, mariano2010coincident, brave2019new, mariano2003new}. Monthly estimates may offer distinct advantages. For policymakers, they provide a more comprehensive snapshot of the economy, enabling more agile and precise interventions, particularly during periods of rapid change.\footnote{\citet{koop2023reconciled} analyze detailed dynamics, such as the timing of economic troughs and the speed of recovery, that cannot be observed using quarterly GDP data. The Office for National Statistics in the United Kingdom evaluates the effects of policy changes such as lockdowns and fiscal support, as well as the recovery patterns across industries, using monthly GDP data \citep{scruton2018introducing}.} For financial market participants, they can serve as a valuable benchmark that reduces uncertainty by synthesizing the noise from various individual indicators. Furthermore, they allow researchers to capture short-run dynamics with greater precision, such as identifying economic turning points more accurately and analyzing the immediate effects of policy shocks.\footnote{Since monthly GDP is often difficult to observe directly, practitioners may rely on proxy indicators. However, such proxies can be biased depending on the structural characteristics of the economy.}

The classical approach to temporal disaggregation, pioneered by \citet{chow1971best} and extended by \citet{fernandez1981methodological} and \citet{litterman1983random}, relies on regression models with autocorrelated residuals to distribute quarterly aggregates across months. These methods are widely used by statistical offices and assume a \textit{linear} relationship between GDP and the monthly indicators. More recently, \citet{koop2023reconciled} developed a Bayesian mixed-frequency vector autoregressive model that integrates quarterly GDP with a range of monthly indicators to estimate monthly GDP in the United States. \citet{chan2023high} proposed an efficient precision-based sampling approach for high-dimensional conditionally Gaussian state space models with missing data, enabling weekly GDP estimation from mixed-frequency data. A common thread across these approaches is that they assume linearity in the mapping from indicators to GDP.

A natural question is whether this linearity assumption is restrictive. Macroeconomic relationships are potentially nonlinear, particularly during periods of large structural change such as the Global Financial Crisis (GFC) or the COVID-19 pandemic. During such episodes, the marginal relationship between indicators and GDP may shift abruptly: a given decline in industrial production, for instance, may be associated with a disproportionately larger contraction in GDP than what a linear model calibrated on tranquil data would predict. Machine learning methods are well-suited to capture such nonlinearities without requiring prior specification of functional form.\footnote{For a discussion on the usefulness of the underlying features driving the gains from machine learning over standard macroeconometric methods, see \citet{goulet2022machine}.} However, the applied literature has not yet systematically evaluated \textit{when} and \textit{why} the flexibility of machine learning methods translates into measurably better monthly GDP estimates relative to classical linear disaggregation.

This paper addresses this gap. We propose a modular framework in which any supervised learning model is trained on quarterly GDP and monthly indicators, then applied to monthly data to generate a high-frequency signal. This signal is reconciled with observed quarterly aggregates using the log-linear approximation of \citet{mariano2003new, mariano2010coincident}. The framework nests the classical \citet{chow1971best} method as a special case and accommodates both linear and nonlinear extensions. We evaluate four models within this framework: the \citet{chow1971best} method (the classical benchmark), Elastic Net \citep{zou2005regularization} (a regularized linear model), XGBoost \citep{chen2016xgboost} (a nonlinear tree-based ensemble), and a Multi-Layer Perceptron \citep[MLP;][]{rumelhart1986learning} (a nonlinear neural network). All models are evaluated under the same data, preprocessing, reconciliation procedure, and expanding-window out-of-sample protocol.\footnote{An earlier version of this paper also considered Long Short-Term Memory networks \citep[LSTM;][]{hochreiter1997long}. After quarterly aggregation, the effective sample size is too small for recurrent architectures to be trained reliably without severe overfitting, so the LSTM is excluded from the current analysis.}

The paper makes three contributions. First, we provide theoretical results characterizing when and how different model features improve temporal disaggregation. We formalize a regime-switching data-generating process (DGP) in which the relationship between GDP and indicators differs across normal and crisis periods, and show that the linear estimator is biased in the crisis regime (\Cref{sec:theory}). Applying the classical result of \citet{hoerl1970ridge}, we show that regularization reduces mean squared error relative to the unpenalized estimator in the high-dimensional indicator environments typical of temporal disaggregation. Second, we provide the first systematic empirical comparison of classical and machine learning temporal disaggregation across four countries using a rigorous out-of-sample evaluation protocol with \citet{diebold1995comparing} tests. Third, we use SHAP values \citep{lundberg2017unified} to identify \textit{when} and through \textit{which variables} model differences arise, providing interpretable evidence on the source of gains.

The evaluation design addresses several methodological concerns. We use an expanding-window out-of-sample protocol rather than a single train-test split, ensuring that performance is assessed across different economic regimes. For the machine learning models, the architecture is selected via Bayesian optimization on the initial training window and held fixed as the window expands; at each subsequent step, only the model weights are re-estimated. The \citet{diebold1995comparing} test with Newey-West heteroskedasticity and autocorrelation consistent (HAC) variance estimation determines whether performance differences are statistically significant.

The framework is applied to four countries with diverse data environments: China, Germany, the United Kingdom, and the United States. For the United States, we benchmark against the monthly GDP estimates of \citet{koop2023reconciled}. For the United Kingdom, we compare against the official monthly GDP series from the Office for National Statistics. For China, we note that the reliability of official macroeconomic statistics has been questioned in the literature.\footnote{See \citet{holz2014quality}, \citet{rawski2001happening}, and \citet{fernald2016reassessing} for discussions of Chinese GDP data quality, including concerns about incomplete seasonal adjustment, potential political manipulation, and inconsistencies between provincial and national aggregates.}

The main findings are as follows. For the United States, the Chow-Lin method performs best with contemporaneous indicators only ($R^2 = 0.72$ at lag~0), but degrades sharply as lagged indicators are added ($R^2 = -1.07$ at lag~2), consistent with the variance inflation predicted by our theoretical analysis. Elastic Net, by contrast, improves with lagged indicators and achieves $R^2 = 0.87$ at lag~1, the strongest result across all configurations. This contrast illustrates the central finding: regularization enables the use of richer information sets that the classical method cannot exploit. Nonlinear models (MLP and XGBoost) do not systematically outperform the linear alternatives in any country, reflecting the bias-variance tradeoff discussed in \Cref{rem:finite}. The framework is applied to four countries with diverse data environments to assess generalizability.

The remainder of the paper is organized as follows. \Cref{sec:methodology} presents the methodology, including the disaggregation framework, the theoretical motivation for nonlinearity, the model specifications, and the evaluation design. \Cref{sec:data} describes the data. \Cref{sec:results} presents the main results. \Cref{sec:conclusion} concludes.

\section{Methodology \label{sec:methodology}}

\subsection{Temporal Disaggregation Framework \label{sub:framework}}

This section describes the framework employed to estimate monthly GDP.\footnote{The algorithms used to implement this framework are detailed in \Cref{app:algorithms}.} Let $Y_q$ denote the observed quarterly real GDP growth (log-difference) for quarter $q$, and let $X_m \in \mathbb{R}^k$ represent the vector of $k$ explanatory variables observed in month $m$. Typical explanatory variables include industrial production, retail sales, the unemployment rate, consumer sentiment, and financial market indicators.\footnote{For example, \citet{koop2023reconciled} use average weekly hours in manufacturing, the consumer price index, industrial production, real personal consumption expenditures, the federal funds rate, the 10-year Treasury rate, the S\&P 500 stock price index, and the civilian unemployment rate.} The objective is to estimate unobserved monthly GDP growth rates $y_m$ that are consistent with the observed quarterly aggregates.

The estimation proceeds in three steps.

\paragraph{Step 1: Quarterly model estimation.} We first aggregate the monthly variables to construct quarterly-level variables $X_q \in \mathbb{R}^{k'}$. For each variable $j$, the quarterly value $X_q^{(j)}$ is constructed as follows:

\begin{enumerate}
    \item For variables measured in levels (e.g., the unemployment rate), we compute the quarterly average:
    \begin{equation}
        X_q^{(j)} = \frac{1}{3} \sum_{m \in q} X_m^{(j)}.
    \end{equation}

    \item For variables expressed as growth rates (e.g., first differences of logarithms), we compute the quarterly sum, consistent with the cumulative growth identity:
    \begin{equation}
        X_q^{(j)} = \sum_{m \in q} X_m^{(j)}.
    \end{equation}
    
    \item For variables measured in rates (e.g., interest rates), where logarithmic transformation is economically inappropriate, we compute the quarterly sum of first differences:
    \begin{equation}
        X_q^{(j)} = \sum_{m \in q} \Delta X_m^{(j)}.
    \end{equation}
\end{enumerate}

In practice, we determine which variables require log-differencing or first-differencing based on the Augmented Dickey-Fuller test \citep{said1984testing} and economic considerations. After constructing the transformed series and aggregating all variables to a quarterly frequency, we split the dataset into training and test sets. To prevent data leakage, we standardize the non-differenced level variables only after this split, applying Z-score standardization using only the training sample moments.\footnote{Each variable $X_j$ is transformed into $z_j = (X_j - \mu_j) / \sigma_j$, where $\mu_j$ and $\sigma_j$ denote the sample mean and standard deviation computed from the training set only.}

We then specify a general regression framework:
\begin{equation} \label{eq:quarterly_model}
    Y_q = f(X_q; \theta) + e_q,
\end{equation}
where $f(\cdot)$ denotes the potentially nonlinear function to be estimated, $\theta$ represents the set of model parameters, and $e_q$ is the error term.

\paragraph{Step 2: Monthly signal generation.} The estimated quarterly model is applied to the original monthly indicators to generate a preliminary monthly signal:
\begin{equation}
    \tilde{y}_m = f(X_m; \hat{\theta}).
\end{equation}
Since $f$ was trained on quarterly data, each $\tilde{y}_m$ reflects the model's assessment of economic conditions in month $m$, but is not directly interpretable as a monthly growth rate. The resulting series $\{\tilde{y}_m\}$ serves as a high-frequency, model-based signal that captures intra-quarter variation as inferred from monthly movements in the explanatory variables.

\paragraph{Step 3: Reconciliation via Mariano-Murasawa constraints.} The preliminary signal is reconciled with observed quarterly GDP using the log-linear approximation of \citet{mariano2003new, mariano2010coincident}. Under this approximation, the quarterly observation $Y_q$ is related to monthly growth rates $y_m$ via a five-term moving average:
\begin{equation} \label{eq:ma5}
    Y_q = \frac{1}{3} y_m + \frac{2}{3} y_{m-1} + y_{m-2} + \frac{2}{3} y_{m-3} + \frac{1}{3} y_{m-4},
\end{equation}
where $m$ denotes the last month of quarter $q$. This relationship can be expressed in matrix form as $My = z$, where $M$ is the constraint matrix encoding the weights in \eqref{eq:ma5} and $z$ is the vector of observed quarterly growth rates. Given the preliminary signal $\tilde{y}$, the reconciled estimates are obtained by solving:
\begin{equation} \label{eq:reconciliation}
    \hat{y} = \tilde{y} + M^\top (M M^\top)^{-1} (z - M\tilde{y}),
\end{equation}
which is the minimum-norm adjustment that satisfies the quarterly constraints exactly.\footnote{The matrix formulation and derivation are provided in \Cref{app:reconciliation}.} This reconciliation replaces the simpler proportional method of \citet{denton1971adjustment}, which assumes $\sum_{m \in q} y_m = Y_q$. The five-term moving average formulation in \eqref{eq:ma5} is the standard approximation in the mixed-frequency literature and is consistent with the treatment in \citet{koop2023reconciled} and \citet{chan2023high}.

\subsection{Theoretical Motivation \label{sec:theory}}

The classical \citet{chow1971best} method assumes a time-invariant linear relationship between GDP growth and monthly indicators. In this section, we formalize conditions under which this assumption leads to biased disaggregation, show that regularization can improve upon the unpenalized estimator, and discuss the finite-sample limitations of nonlinear alternatives.

\begin{proposition}[Bias of linear disaggregation under regime switching] \label{prop:bias}
Suppose the true data-generating process is regime-dependent:
\begin{equation} \label{eq:dgp_regime}
    Y_q = \begin{cases}
        \beta_1^\top X_q + \varepsilon_q & \text{if } s_q = 0 \quad \text{(normal)}, \\
        \beta_2^\top X_q + \varepsilon_q & \text{if } s_q = 1 \quad \text{(crisis)},
    \end{cases}
\end{equation}
where $s_q \in \{0, 1\}$ is an unobserved regime indicator, $\varepsilon_q$ is i.i.d.\ with mean zero, and $\beta_1 \neq \beta_2$. Let $\pi = \Pr(s_q = 1)$ denote the unconditional probability of the crisis regime. Then:

\begin{enumerate}
    \item[(i)] The ordinary least squares estimator $\hat{\beta}_{\text{OLS}}$ converges in probability to
    \begin{equation}
        \bar{\beta} = (1 - \pi)\,\beta_1 + \pi\,\beta_2,
    \end{equation}
    a weighted average of the regime-specific coefficients.
    
    \item[(ii)] In the crisis regime ($s_q = 1$), the linear estimator's prediction error has conditional bias
    \begin{equation}
        \mathbb{E}[Y_q - \bar{\beta}^\top X_q \mid s_q = 1] = (1 - \pi)(\beta_2 - \beta_1)^\top \mathbb{E}[X_q \mid s_q = 1],
    \end{equation}
    which is nonzero whenever $\beta_1 \neq \beta_2$.
    
    \item[(iii)] If crises are rare ($\pi \to 0$), the bias in the crisis regime approaches $(\beta_2 - \beta_1)^\top \mathbb{E}[X_q \mid s_q = 1]$, the full misspecification gap.
\end{enumerate}
\end{proposition}

The proof is provided in \Cref{app:proofs}. The DGP in \eqref{eq:dgp_regime} is a special case of the threshold regression framework of \citet{hansen2000sample}. The key insight is not that the linear model performs poorly on average, but that it performs poorly \textit{precisely when accuracy matters most}: during crises. When crises are rare, the linear model is well-calibrated for normal times but systematically underpredicts the severity of contractions and the speed of recoveries.

A natural response is to consider nonlinear estimators. By the universal approximation theorem \citep{hornik1989multilayer}, a feedforward neural network with a single hidden layer of sufficient width can approximate any continuous function on a compact set to arbitrary precision, including the piecewise-linear mapping in \eqref{eq:dgp_regime}. However, approximation capacity does not guarantee estimation quality in finite samples. An alternative approach is to retain linearity but introduce regularization to improve estimation precision, particularly when the number of indicators is large relative to the sample size. We apply the classical result of \citet{hoerl1970ridge} to the temporal disaggregation context:

\begin{proposition}[MSE reduction via regularization] \label{prop:ridge}
Consider the linear model $Y = X\beta + \varepsilon$ with $\varepsilon \sim (0, \sigma^2 I_n)$, where $X \in \mathbb{R}^{n \times k}$ has full column rank. Let $\hat{\beta}_{\text{OLS}} = (X^\top X)^{-1} X^\top Y$ and let $\hat{\beta}_\lambda = (X^\top X + \lambda I_k)^{-1} X^\top Y$ for $\lambda > 0$. Then there exists $\lambda^* > 0$ such that
\begin{equation}
    \text{MSE}(\hat{\beta}_{\lambda^*}) < \text{MSE}(\hat{\beta}_{\text{OLS}}),
\end{equation}
where $\text{MSE}(\hat{\beta}) = \mathbb{E}\left[\|\hat{\beta} - \beta\|^2\right]$.
\end{proposition}

The proof, which follows directly from \citet{hoerl1970ridge}, is provided in \Cref{app:proofs}. In the context of temporal disaggregation, this result is relevant when the number of monthly indicators (including their lags) grows relative to the number of observed quarters. The \citet{chow1971best} estimator, being an unpenalized GLS estimator, is susceptible to the high variance associated with near-collinear regressors. Elastic Net, which combines $\ell_1$ and $\ell_2$ penalties, inherits the MSE improvement of \Cref{prop:ridge} while additionally performing variable selection through the $\ell_1$ component.

\begin{remark}[Finite-sample tradeoff for nonlinear estimators] \label{rem:finite}
\Cref{prop:bias} establishes that a flexible estimator can eliminate the conditional bias of linear disaggregation under regime switching. However, this is an asymptotic result. In finite samples, the variance of a nonlinear estimator with $p$ parameters estimated from $n$ observations may exceed the bias reduction, yielding higher mean squared error than a regularized linear estimator. When $n$ is small relative to $p$, as is typical with quarterly GDP data (60--130 observations), regularization of a linear model can achieve a more favorable bias-variance tradeoff than a nonlinear model. The empirical analysis in \Cref{sec:results} examines this tradeoff directly.
\end{remark}

\subsection{Models \label{sub:models}}

We evaluate four models that span a range of flexibility and regularization within the general framework of \eqref{eq:quarterly_model}.

\paragraph{Chow-Lin (classical benchmark).}

The classical \citet{chow1971best} method serves as the primary benchmark. It estimates the coefficient vector $\beta$ by generalized least squares (GLS), assuming that the residuals at the monthly frequency follow a first-order autoregressive process with parameter $\rho$. The optimal $\rho$ is found by maximizing the GLS log-likelihood. Given the estimated $\hat{\beta}$ and $\hat{\rho}$, monthly GDP is distributed using the estimated relationship between GDP and the monthly indicators, with the AR covariance structure governing how the quarterly residual is allocated across months. This method is widely used by statistical offices and represents the standard approach to temporal disaggregation.\footnote{Detailed mathematical specifications are provided in \Cref{app:models}.}

\paragraph{Elastic Net (regularized linear).}

The Elastic Net \citep{zou2005regularization} extends the linear framework with $\ell_1$ and $\ell_2$ penalties, combining the variable selection property of the LASSO with the grouping effect of Ridge regression. In the context of \Cref{prop:ridge}, the Elastic Net inherits the MSE improvement of ridge-type regularization while additionally shrinking irrelevant coefficients to zero. This is particularly relevant when the number of monthly indicators and their lags approaches or exceeds the number of quarterly observations, a setting where the unpenalized \citet{chow1971best} estimator suffers from high variance. The regularization parameters are selected via cross-validation on the training window.\footnote{Detailed specifications, including the bootstrap inference procedure, are provided in \Cref{app:models}.}

\paragraph{XGBoost (nonlinear, tree-based).}

XGBoost \citep{chen2016xgboost} is a gradient boosting ensemble of decision trees. Unlike the linear models, XGBoost can capture nonlinear relationships and interactions between indicators without explicit specification. Key hyperparameters are optimized via cross-validated grid search.\footnote{See \Cref{app:models} for the full specification and hyperparameter grid.}

\paragraph{Multi-Layer Perceptron (MLP; nonlinear, neural network).}

The Multi-Layer Perceptron is a class of feedforward artificial neural networks capable of capturing complex nonlinear relationships \citep{rumelhart1986learning}. An MLP models the target variable as a nested series of nonlinear transformations of the explanatory variables. For a network with $L$ hidden layers, the output is computed recursively:
\begin{align}
    h^{(0)} &= X_q, \notag \\
    h^{(\ell)} &= \sigma^{(\ell)}\!\left( W^{(\ell)} h^{(\ell-1)} + b^{(\ell)} \right), \quad \ell = 1, \ldots, L, \\
    f(X_q; \theta) &= W^{(L+1)} h^{(L)} + b^{(L+1)}, \notag
\end{align}
where $W^{(\ell)}$ and $b^{(\ell)}$ are the weight matrix and bias vector for layer $\ell$, $\sigma^{(\ell)}$ is the activation function, and $\theta = \{W^{(\ell)}, b^{(\ell)}\}_{\ell=1}^{L+1}$ is the full set of parameters. The model parameters are estimated by minimizing the mean squared error using the Adam optimizer \citep{kingma2014adam}. The architecture, including the number of hidden layers (up to 2), neurons per layer (up to 128), activation functions, and dropout rates, is selected via Bayesian hyperparameter optimization on the initial training window.\footnote{We utilize the \texttt{Keras-Tuner} package. The search space is constrained to shallow architectures to mitigate overfitting given the small effective sample size after quarterly aggregation. See \Cref{app:algorithms} for the complete training algorithm.}

In the context of \Cref{prop:bias}, the MLP replaces the regime-invariant linear function $\bar{\beta}^\top X_q$ with a flexible nonlinear mapping $f(X_q; \theta)$ that can adapt to different regimes without requiring explicit regime identification. However, as noted in \Cref{rem:finite}, the estimation variance of the MLP may offset this theoretical advantage when the sample size is small. To mitigate overfitting, three safeguards are employed: the search space is constrained to shallow architectures (up to 2 layers, 128 neurons); the architecture is selected only on the initial training window and held fixed for subsequent windows; and the validation set used for early stopping and model selection is drawn from the training window (last 20\%), never from the test data.

\subsection{Evaluation Design \label{sub:evaluation}}

\paragraph{Expanding-window out-of-sample evaluation.} Rather than relying on a single train-test split, we employ an expanding-window protocol. The initial training window covers 50\% of the available sample, ensuring that the test period includes both the Global Financial Crisis and the COVID-19 pandemic. At each step, the model is retrained on all available data up to period $t$, a one-quarter-ahead prediction is generated, and the window expands by one quarter. This procedure yields a sequence of genuine out-of-sample predictions across different economic regimes. For the MLP and XGBoost, the architecture or hyperparameters are selected on the initial training window and held fixed as the window expands; at each subsequent step, only the model weights are re-estimated on the expanded training data. This is consistent with standard practice in the forecasting literature, where model specification is not revised at each evaluation step.

\paragraph{Diebold-Mariano tests.} To assess whether performance differences between models are statistically significant, we apply the \citet{diebold1995comparing} test with Newey-West HAC variance estimation. All pairwise comparisons are reported in \Cref{app:additional}.

\paragraph{Accuracy metrics.} We report root mean squared error (RMSE), mean absolute error (MAE), $R^2$, correlation, and sign accuracy for the quarterly prediction task. For the monthly estimates, we compare against the benchmark series of \citet{koop2023reconciled} for the United States and the official series from the Office for National Statistics for the United Kingdom.

\paragraph{SHAP-based interpretability.} To understand \textit{when} and \textit{why} models differ in their predictions, we compute SHAP values \citep{lundberg2017unified} for all models. SHAP provides a unified, theoretically grounded measure of each variable's contribution to each prediction, enabling direct comparison of variable importance across linear and nonlinear models.\footnote{XGBoost's built-in importance scores are based on gain, while Elastic Net coefficients reflect marginal effects under a linear assumption. SHAP values, by contrast, are model-agnostic and based on cooperative game theory, making them comparable across arbitrary model classes.}

\paragraph{Reconciliation diagnostics.} We report the implicit adjustment factor for each quarter, defined as the ratio of the observed quarterly growth to the model's pre-reconciliation quarterly aggregate. Values close to unity indicate that the model's raw monthly signal is already well-calibrated. Large variation in the adjustment factor indicates that the reconciliation step is doing substantial work, which may signal model misspecification at the monthly frequency.

\section{Data \label{sec:data}}

For the empirical analysis, we collect data for four countries: China, Germany, the United Kingdom, and the United States. While our target sample spans from 1991 to 2024, the actual coverage varies by country depending on the availability of data. We cap all samples at 2024Q4 to ensure that only finalized (non-preliminary) data are used; China's sample ends earlier (2023Q3) due to data availability constraints. In all cases, we ensure that the quarterly GDP series and all explanatory variables are temporally aligned within a consistent time frame for each country.

Quarterly real GDP growth is calculated as the log-difference of seasonally adjusted real GDP. The monthly explanatory variables are organized into seven economic categories: production, expenditure, financial markets, labor, trade, monetary aggregates, and prices. This categorization is intended to reflect the key dimensions of GDP measurement and to provide transparency regarding indicator selection.\footnote{The choice of indicator categories follows the structure of GDP measurement in each country. For the United States, where GDP is measured primarily from the expenditure side, indicators such as real personal consumption expenditure and net exports play a central role. For Germany and the United Kingdom, where the production approach is more prominent, industrial production and manufacturing output receive greater weight.} The specific variables within each category vary across countries depending on data availability. Table~\ref{tab:sum_data} summarizes the datasets used for each country, with detailed descriptions provided in \Cref{app:data}.

\begin{table}[htbp!]
\begin{center}
\caption{Summary of Datasets}
\begin{tabular}{lcccc}
\toprule
Variable & China & Germany & UK & US \\
\midrule
\multicolumn{5}{l}{\textit{Production}} \\
Industrial Production             &        &        &      & FRED \\
Production Volume                 &        & FRED   & FRED &      \\
Avg.\ Weekly Hours                &        &        &      & FRED \\

\multicolumn{5}{l}{\textit{Expenditure}} \\
Real Personal Consum.             &        &        &      & FRED \\
Retail Trade                      &        & FRED   &      &      \\

\multicolumn{5}{l}{\textit{Financial}} \\
S\&P / FTSE / DAX / SSEC         &  Invest    & YF     & YF   & YF   \\
S\&P 500 Volatility               &        &        &      & FRED \\
Share Prices                      &        &        & FRED &      \\
Interest Rate (10Y)               &        &        & FRED & FRED \\
Interest Rate (Gov Bond)          &        &        & FRED &      \\
Effect.\ Fund Rate                &        &        &      & FRED \\
Moody's AAA Bond                  &        &        &      & FRED \\

\multicolumn{5}{l}{\textit{Labor}} \\
Unemployment Rate                 &        & FRED   & ONS  & FRED \\
Employment Rate                   &        &        & ONS  &      \\
Labor Force Participation         &        &        &      & FRED \\
Nonfarm Payroll Emp.              &        &        &      & FRED \\

\multicolumn{5}{l}{\textit{Trade}} \\
(Net) Exports                     & FRED   & FRED   & FRED & FRED \\
Imports                           & FRED   & FRED   & FRED &      \\
US Imports from China             & FRED   &        &      &      \\
Effect.\ Ex.\ Rate               & FRED   & FRED   &      &      \\

\multicolumn{5}{l}{\textit{Monetary}} \\
M1 Money Stock                    &        &        & BOE  & FRED \\
M2 Money Stock                    &        &        &      & FRED \\
Total Reserves                    & FRED   & FRED   & FRED &      \\
Exchange Rate                     & FRED   & FRED   & FRED &      \\

\multicolumn{5}{l}{\textit{Prices}} \\
CPI                               & FRED   & FRED   & FRED & FRED \\
PPI                               & FRED   &        &      &      \\
Price Competitiv.                 &        & DB     &      &      \\
\addlinespace
\multicolumn{5}{l}{\textit{Other}} \\
Policy Uncertainty                & FRED   & FRED   &      &      \\
\midrule
Total Variables Used              & 10     & 12     & 13   & 15   \\
Data Span                         & 01.94--09.23 & 01.91--03.24 & 01.91--03.24 & 01.92--12.24 \\
\bottomrule
\end{tabular}
\label{tab:sum_data}
\vspace{0.3em}
\begin{minipage}{1\textwidth}
{\footnotesize \textit{Notes}: The entries indicate the primary data source. FRED: Federal Reserve Economic Data; YF: Yahoo Finance; ONS: Office for National Statistics; BOE: Bank of England; DB: Deutsche Bundesbank; Invest: Investing.com. Variable counts exclude quarterly real GDP.}
\end{minipage}
\end{center}
\end{table}

The variables are collected from central banks, capital market authorities, national statistical offices, and other third-party providers. Most datasets are retrieved via the FRED API (Federal Reserve Economic Data), with additional series obtained from Yahoo Finance, the Bank of England, the Office for National Statistics, the Deutsche Bundesbank, and Investing.com.\footnote{Detailed instructions for data acquisition, including Python scripts for API-based access, are provided in the replication package.}

For each variable, we determine the appropriate transformation based on the Augmented Dickey-Fuller test \citep{said1984testing} and economic considerations. Variables representing price levels, quantity indices, and stock market indices are log-differenced. Interest rates, which are already expressed in percentage terms, are first-differenced without logarithmic transformation. Variables that are stationary in levels, such as the unemployment rate in some countries, are retained without differencing and instead standardized.

For China, where officially published GDP and some monthly indicators may not be fully seasonally adjusted, we apply Seasonal-Trend decomposition using LOESS \citep[STL;][]{cleveland1990stl} to all explanatory variables prior to model estimation. This addresses concerns about residual seasonality in the Chinese data.

For the United States, monthly GDP growth estimates produced by \citet{koop2023reconciled} are available, while the Office for National Statistics in the United Kingdom publishes official monthly GDP figures.\footnote{Available at \href{https://www.ons.gov.uk/economy/grossdomesticproductgdp}{https://www.ons.gov.uk/economy/grossdomesticproductgdp}.} These two cases allow us to benchmark the performance of our framework against established references. For the remaining countries, we demonstrate that the framework produces plausible estimates and examine its robustness across heterogeneous data environments.

Detailed variable definitions, FRED series codes, summary statistics, and Augmented Dickey-Fuller test results are provided in \Cref{app:data}. All data collection and processing code is available in the replication package.

\section{Results \label{sec:results}}

\subsection{Quarterly Prediction Performance \label{sub:quarterly}}

Table~\ref{tab:main_results} reports the expanding-window out-of-sample performance of all four models at lag~0 (contemporaneous indicators only) across the four countries. The United States stands out as the only country where models achieve substantial predictive power: the Chow-Lin benchmark attains $R^2 = 0.72$ and correlation 0.87, with the MLP also performing reasonably ($R^2 = 0.41$). For Germany, the United Kingdom, and China, all models produce $R^2$ values below 0.23, indicating that contemporaneous indicators alone explain little of the quarterly variation in GDP growth in these countries. This is not a model failure per se; rather, it reflects differences in data quality, the number of available indicators, and the degree to which monthly indicators track GDP movements.

\begin{table}[htbp]
\centering
\caption{Quarterly Prediction Performance: Expanding-Window Out-of-Sample Results (Lag = 0)}
\label{tab:main_results}
\begin{tabular}{ll ccccc}
\toprule
Country & Model & RMSE & MAE & $R^2$ & Correlation & Sign Acc. \\
\midrule
United States & Chow-Lin & \textbf{0.0085} & 0.0053 & 0.723 & 0.866 & 92.5\% \\
 & Elastic Net & 0.0115 & 0.0061 & 0.488 & 0.785 & 96.2\% \\
 & MLP & 0.0123 & 0.0069 & 0.414 & 0.657 & 96.2\% \\
 & XGBoost & 0.0153 & 0.0063 & 0.103 & 0.370 & 94.3\% \\
\addlinespace
Germany & Elastic Net & \textbf{0.0160} & 0.0068 & 0.225 & 0.521 & 74.1\% \\
 & MLP & 0.0165 & 0.0075 & 0.168 & 0.410 & 64.8\% \\
 & Chow-Lin & 0.0165 & 0.0076 & 0.167 & 0.410 & 77.8\% \\
 & XGBoost & 0.0170 & 0.0076 & 0.124 & 0.375 & 72.2\% \\
\addlinespace
United Kingdom & MLP & \textbf{0.0368} & 0.0135 & 0.092 & 0.314 & 83.3\% \\
 & Chow-Lin & 0.0375 & 0.0115 & 0.058 & 0.272 & 74.1\% \\
 & Elastic Net & 0.0378 & 0.0119 & 0.042 & 0.232 & 79.6\% \\
 & XGBoost & 0.0381 & 0.0120 & 0.029 & 0.187 & 83.3\% \\
\addlinespace
China & Elastic Net & \textbf{0.1143} & 0.0939 & 0.078 & 0.457 & 79.2\% \\
 & Chow-Lin & 0.1143 & 0.0967 & 0.077 & 0.278 & 77.1\% \\
 & XGBoost & 0.1160 & 0.0930 & 0.049 & 0.287 & 75.0\% \\
 & MLP & 0.1300 & 0.1075 & $-$0.193 & 0.078 & 62.5\% \\
\bottomrule
\end{tabular}
\vspace{0.3em}
\begin{minipage}{1\textwidth}
\footnotesize \textit{Notes:} Bold RMSE indicates the best-performing model for each country. The initial training window covers 50\% of the sample. All metrics are computed on genuine out-of-sample predictions.
\end{minipage}
\end{table}

The picture changes substantially when lagged indicators are introduced. Table~\ref{tab:us_lags} reports the United States results across lag specifications 0, 1, and 2. Three patterns emerge. First, the Chow-Lin method degrades sharply as the number of regressors grows: its $R^2$ drops from 0.72 at lag~0 to $-1.07$ at lag~2, consistent with the variance inflation predicted when near-collinear regressors are estimated without regularization. Second, Elastic Net improves with lagged indicators, achieving its best performance at lag~1 ($R^2 = 0.87$, correlation 0.94). The $\ell_1$ penalty effectively selects the most informative lagged indicators while shrinking the rest, confirming the mechanism formalized in \Cref{prop:ridge}. Third, the nonlinear models (MLP and XGBoost) do not benefit from additional lags and remain inferior to Elastic Net at every lag specification. This is consistent with the finite-sample qualification in \Cref{rem:finite}: the variance cost of estimating a flexible nonlinear function from 66 quarterly observations exceeds the bias reduction from capturing nonlinearities.

\begin{table}[htbp]
\centering
\caption{United States: Effect of Lagged Indicators on Model Performance}
\label{tab:us_lags}
\begin{tabular}{cl ccccc}
\toprule
Lag & Model & RMSE & MAE & $R^2$ & Correlation & Sign Acc. \\
\midrule
0 & Chow-Lin & \textbf{0.0085} & 0.0053 & 0.723 & 0.866 & 92.5\% \\
 & Elastic Net & 0.0115 & 0.0061 & 0.488 & 0.785 & 96.2\% \\
 & MLP & 0.0123 & 0.0069 & 0.414 & 0.657 & 96.2\% \\
 & XGBoost & 0.0153 & 0.0063 & 0.103 & 0.370 & 94.3\% \\
\addlinespace
1 & Elastic Net & \textbf{0.0058} & 0.0040 & 0.870 & 0.942 & 96.2\% \\
 & Chow-Lin & 0.0138 & 0.0064 & 0.271 & 0.527 & 83.0\% \\
 & XGBoost & 0.0155 & 0.0065 & 0.074 & 0.298 & 94.3\% \\
 & MLP & 0.0255 & 0.0167 & $-$1.496 & 0.487 & 75.5\% \\
\addlinespace
2 & Elastic Net & \textbf{0.0078} & 0.0050 & 0.771 & 0.897 & 96.2\% \\
 & XGBoost & 0.0151 & 0.0062 & 0.138 & 0.431 & 94.2\% \\
 & MLP & 0.0169 & 0.0074 & $-$0.085 & -0.058 & 88.5\% \\
 & Chow-Lin & 0.0234 & 0.0098 & $-$1.069 & 0.365 & 86.5\% \\
\bottomrule
\end{tabular}
\vspace{0.3em}
\begin{minipage}{1\textwidth}
\footnotesize \textit{Notes:} Lag $k$ means $k$ quarterly lags of all indicators are added to the regressor set. Bold RMSE indicates the best model at each lag. Chow-Lin performance degrades sharply with lags due to multicollinearity, while Elastic Net maintains strong performance through $\ell_1$-regularization.
\end{minipage}
\end{table}

Table~\ref{tab:best_config} summarizes the best model-lag configuration for each country. For the United States, Elastic Net at lag~1 dominates. For Germany, the MLP at lag~1 achieves $R^2 = 0.31$, the only case outside the United States where a model attains meaningful predictive power, and the only case where a nonlinear model is the best performer. For the United Kingdom and China, the best achievable $R^2$ remains below 0.10 regardless of model or lag specification.

\begin{table}[htbp]
\centering
\caption{Best Model-Lag Configuration per Country}
\label{tab:best_config}
\begin{tabular}{l l c cccc}
\toprule
Country & Best Model & Lag & RMSE & $R^2$ & Correlation & Sign Acc. \\
\midrule
United States & Elastic Net & 1 & 0.0058 & 0.870 & 0.942 & 96.2\% \\
Germany & MLP & 1 & 0.0150 & 0.313 & 0.567 & 74.1\% \\
United Kingdom & MLP & 0 & 0.0368 & 0.092 & 0.314 & 83.3\% \\
China & Elastic Net & 0 & 0.1143 & 0.078 & 0.457 & 79.2\% \\
\bottomrule
\end{tabular}
\end{table}

Table~\ref{tab:dm_tests} reports the Diebold-Mariano test results for all pairwise comparisons that are statistically significant at the 10\% level. Of the 72 pairwise tests conducted, 12 are significant at 10\% and 7 at 5\%. The most notable results are: (i) for the United States at lag~1, Chow-Lin and Elastic Net both significantly outperform the MLP at the 1\% level; (ii) for China at lag~2, all three machine learning models significantly outperform Chow-Lin, driven by the latter's variance explosion in the high-dimensional regressor environment; and (iii) for Germany at lag~0, Elastic Net significantly outperforms Chow-Lin at the 10\% level.

It is worth noting that the Elastic Net vs.\ Chow-Lin comparison at lag~1 for the United States, despite the large difference in $R^2$ (0.87 vs.\ 0.27), does not reach statistical significance in the DM test ($p = 0.17$). This reflects the limited power of the test with approximately 53 out-of-sample observations, compounded by the high variance of squared forecast errors during the COVID-19 period. The economic magnitude of the improvement is nonetheless substantial: Elastic Net reduces RMSE by 58\% relative to Chow-Lin at this lag specification.

\begin{table}[htbp]
\centering
\caption{Diebold-Mariano Test Results: Significant Pairwise Comparisons}
\label{tab:dm_tests}
\begin{tabular}{cl cc cl}
\toprule
Country & Lag & Model 1 vs Model 2 & DM Stat & $p$-value & Favors \\
\midrule
China & 2 & Chow-Lin vs XGBoost & +2.364 & 0.0181$^{**}$ & XGBoost \\
China & 2 & Chow-Lin vs MLP & +2.322 & 0.0202$^{**}$ & MLP \\
China & 2 & Chow-Lin vs Elastic Net & +2.081 & 0.0374$^{**}$ & Elastic Net \\
Germany & 0 & Chow-Lin vs Elastic Net & +1.798 & 0.0721$^{*}$ & Elastic Net \\
Germany & 2 & XGBoost vs MLP & -2.549 & 0.0108$^{**}$ & XGBoost \\
Germany & 2 & Chow-Lin vs MLP & -1.946 & 0.0516$^{*}$ & Chow-Lin \\
UK & 2 & Chow-Lin vs MLP & -1.668 & 0.0954$^{*}$ & Chow-Lin \\
UK & 2 & XGBoost vs MLP & -1.668 & 0.0954$^{*}$ & XGBoost \\
US & 1 & Chow-Lin vs MLP & -3.498 & 0.0005$^{***}$ & Chow-Lin \\
US & 1 & Elastic Net vs MLP & -3.155 & 0.0016$^{***}$ & Elastic Net \\
US & 1 & XGBoost vs MLP & -2.545 & 0.0109$^{**}$ & XGBoost \\
US & 2 & XGBoost vs MLP & -1.674 & 0.0941$^{*}$ & XGBoost \\
\bottomrule
\end{tabular}
\vspace{0.3em}
\begin{minipage}{1\textwidth}
\footnotesize \textit{Notes:} Diebold-Mariano test with Newey-West HAC variance. Squared error loss. $^{*}$, $^{**}$, $^{***}$ denote significance at the 10\%, 5\%, and 1\% levels, respectively. Only comparisons significant at the 10\% level are shown. Full pairwise results are in \Cref{app:additional}.
\end{minipage}
\end{table}

\subsection{Monthly GDP Estimates \label{sub:monthly}}

\begin{figure}[htbp]
    \centering
    \includegraphics[width=\textwidth]{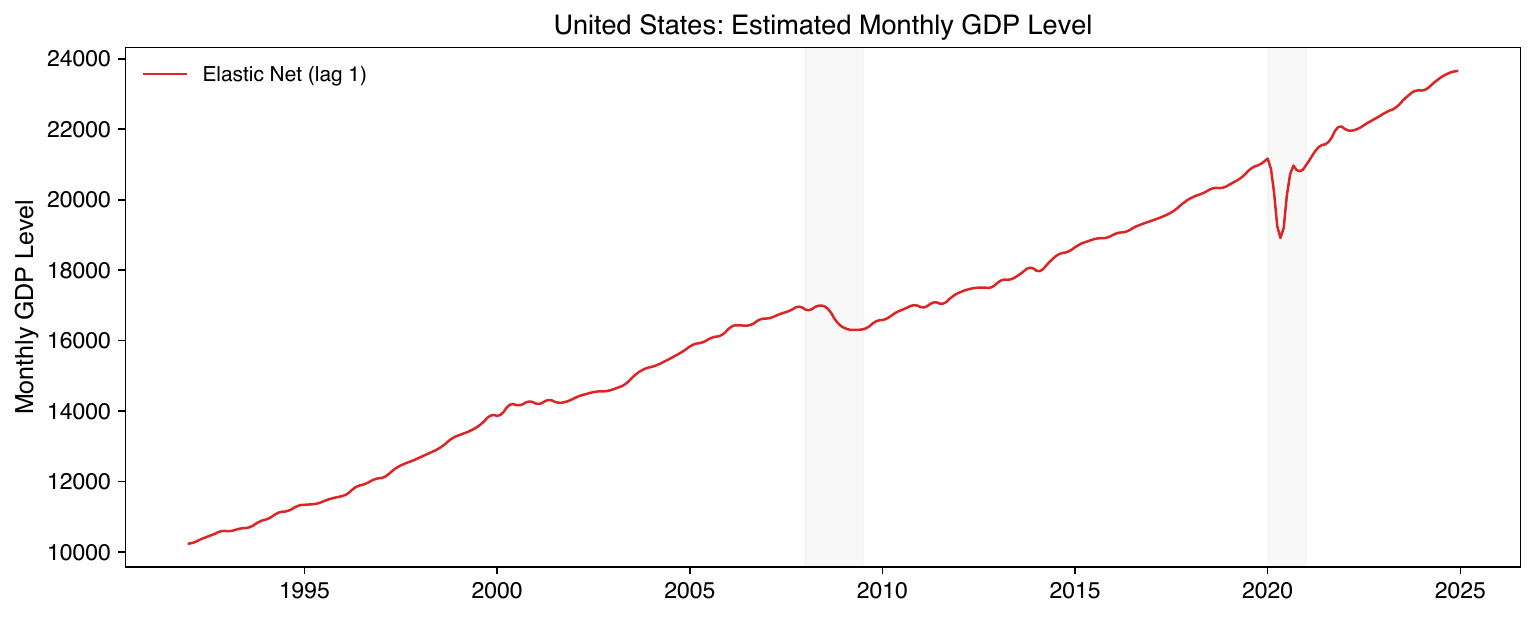}
    \caption{United States: Estimated Monthly GDP Level}
    \label{fig:us_monthly_level}
    \vspace{0.3em}
    \begin{minipage}{1\textwidth}
    {\footnotesize \textit{Notes}: Elastic Net (lag~1). Shaded areas indicate the GFC (2008--2009) and COVID-19 (2020) periods.}
    \end{minipage}
\end{figure}

\begin{figure}[htbp]
    \centering
    \includegraphics[width=\textwidth]{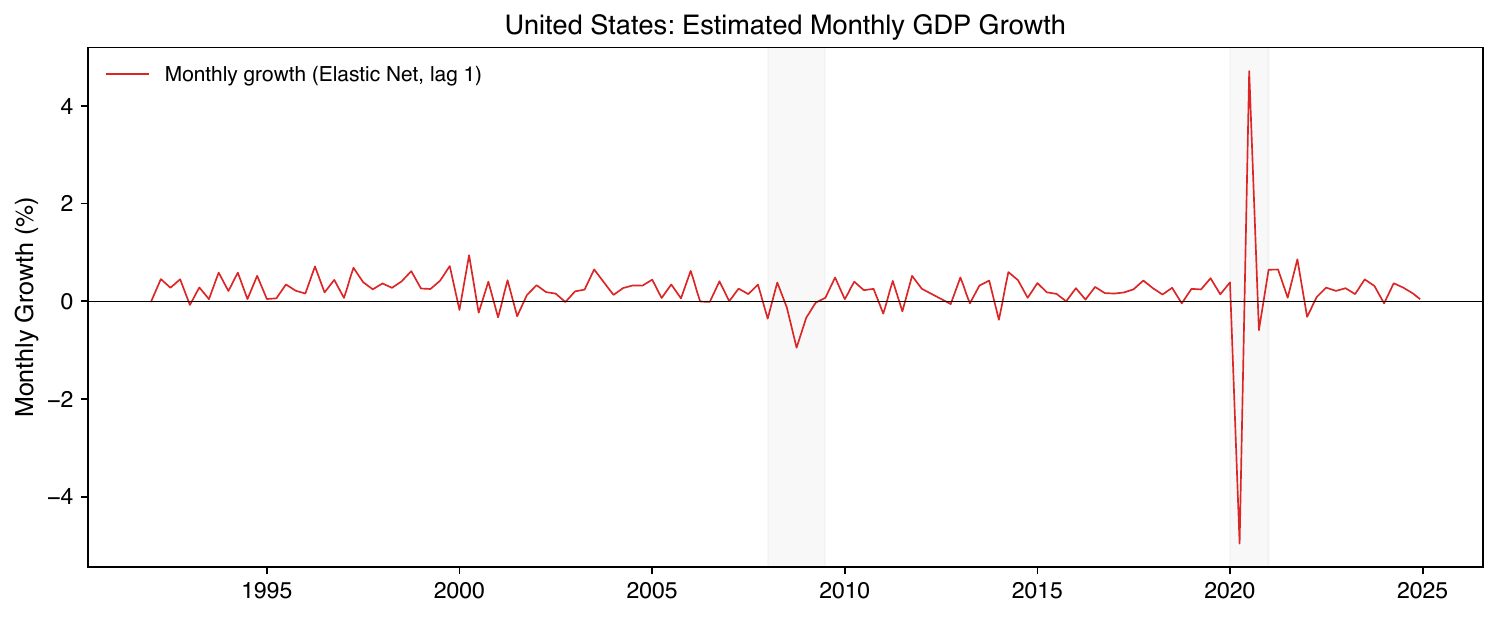}
    \caption{United States: Estimated Monthly GDP Growth}
    \label{fig:us_monthly_growth}
    \vspace{0.3em}
    \begin{minipage}{1\textwidth}
    {\footnotesize \textit{Notes}: Elastic Net (lag~1). Shaded areas indicate the GFC and COVID-19 periods.}
    \end{minipage}
\end{figure}

\Cref{fig:us_monthly_level} displays the estimated monthly GDP level for the United States from the Elastic Net model at lag~1, the best-performing configuration. The series tracks the broad trajectory of the U.S.\ economy, including the contraction during the GFC, the subsequent recovery, and the sharp decline and rebound associated with the COVID-19 pandemic. \Cref{fig:us_monthly_growth} shows the corresponding monthly growth rates, which exhibit the expected pattern of moderate fluctuations during tranquil periods and extreme movements during crises.

\begin{figure}[htbp]
    \centering
    \includegraphics[width=\textwidth]{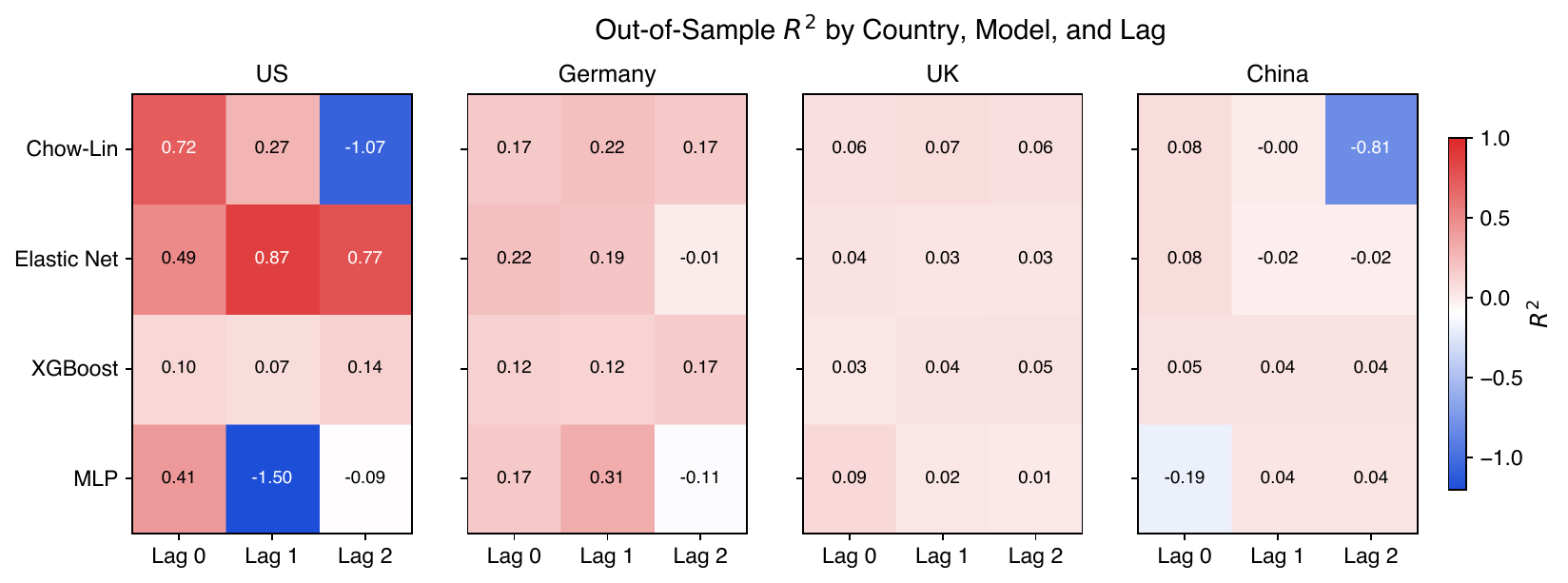}
    \caption{Out-of-Sample $R^2$ by Country, Model, and Lag}
    \label{fig:r2_heatmap}
    \vspace{0.3em}
    \begin{minipage}{1\textwidth}
    {\footnotesize \textit{Notes}: Each cell reports the expanding-window out-of-sample $R^2$. Blue indicates positive $R^2$; red indicates negative $R^2$ (worse than a constant prediction).}
    \end{minipage}
\end{figure}

\begin{figure}[htbp]
    \centering
    \includegraphics[width=0.75\textwidth]{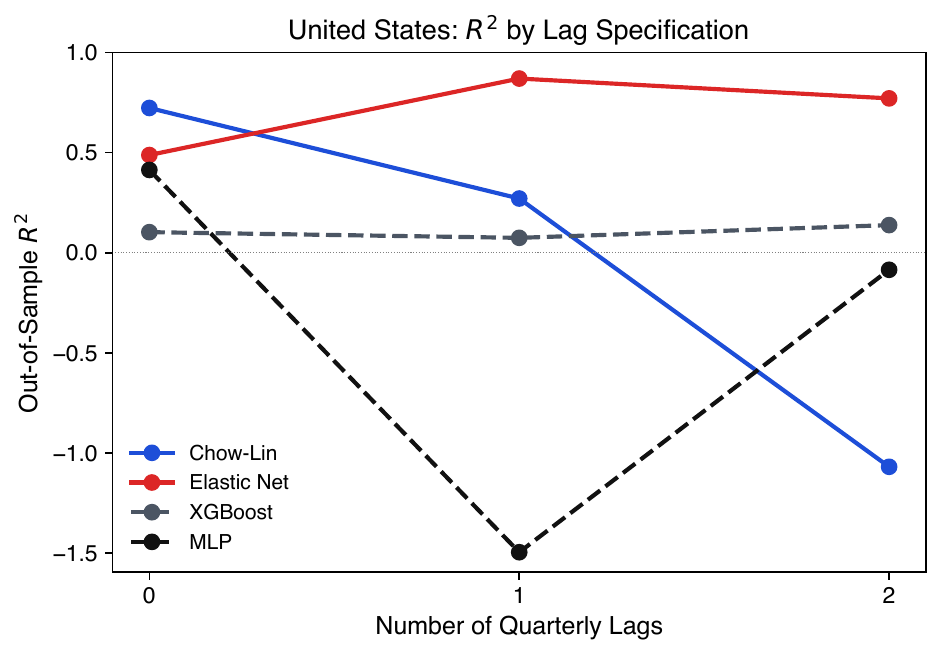}
    \caption{United States: $R^2$ by Lag Specification}
    \label{fig:us_lag_degradation}
    \vspace{0.3em}
    \begin{minipage}{1\textwidth}
    {\footnotesize \textit{Notes}: Expanding-window out-of-sample $R^2$ as a function of the number of quarterly lags added to the regressor set. Chow-Lin degrades sharply due to multicollinearity; Elastic Net improves through $\ell_1$-regularization.}
    \end{minipage}
\end{figure}

\Cref{fig:r2_heatmap} provides a comprehensive overview of out-of-sample $R^2$ across all countries, models, and lag specifications. The figure highlights two key patterns. First, the United States is the only country where models achieve substantial predictive power, with Elastic Net maintaining high $R^2$ across all lag specifications while Chow-Lin degrades sharply. Second, for Germany, the United Kingdom, and China, all models produce $R^2$ values near zero, reflecting the difficulty of predicting GDP growth from monthly indicators in these data environments. \Cref{fig:us_lag_degradation} isolates the United States results, showing the contrasting trajectories of Chow-Lin (which collapses with additional lags) and Elastic Net (which improves).

\subsection{Benchmark Comparisons \label{sub:benchmark}}

\subsubsection{United States: Comparison with \citet{koop2023reconciled}}

\citet{koop2023reconciled} construct monthly GDP estimates by embedding a Bayesian mixed-frequency vector autoregressive model that incorporates quarterly expenditure-side and income-side GDP, a latent true GDP process, and multiple monthly indicators of economic activity. Following their approach, and building on the methodology of \citet{mariano2003new, mariano2010coincident} and \citet{mitchell2005indicator}, we transform our final monthly growth rate $\hat{y}_m$ into an annualized rate for comparison using the same five-month weighted average formula:
\begin{equation}
    \hat{y}^A_{m,t} = \left(\frac{1}{3}\hat{y}_{m,t} + \frac{2}{3}\hat{y}_{m,t-1}+\hat{y}_{m,t-2}+\frac{2}{3}\hat{y}_{m,t-3}+\frac{1}{3}\hat{y}_{m,t-4}\right)\times 4 \times 100.
\end{equation}

\begin{figure}[htbp]
    \centering
    \includegraphics[width=\textwidth]{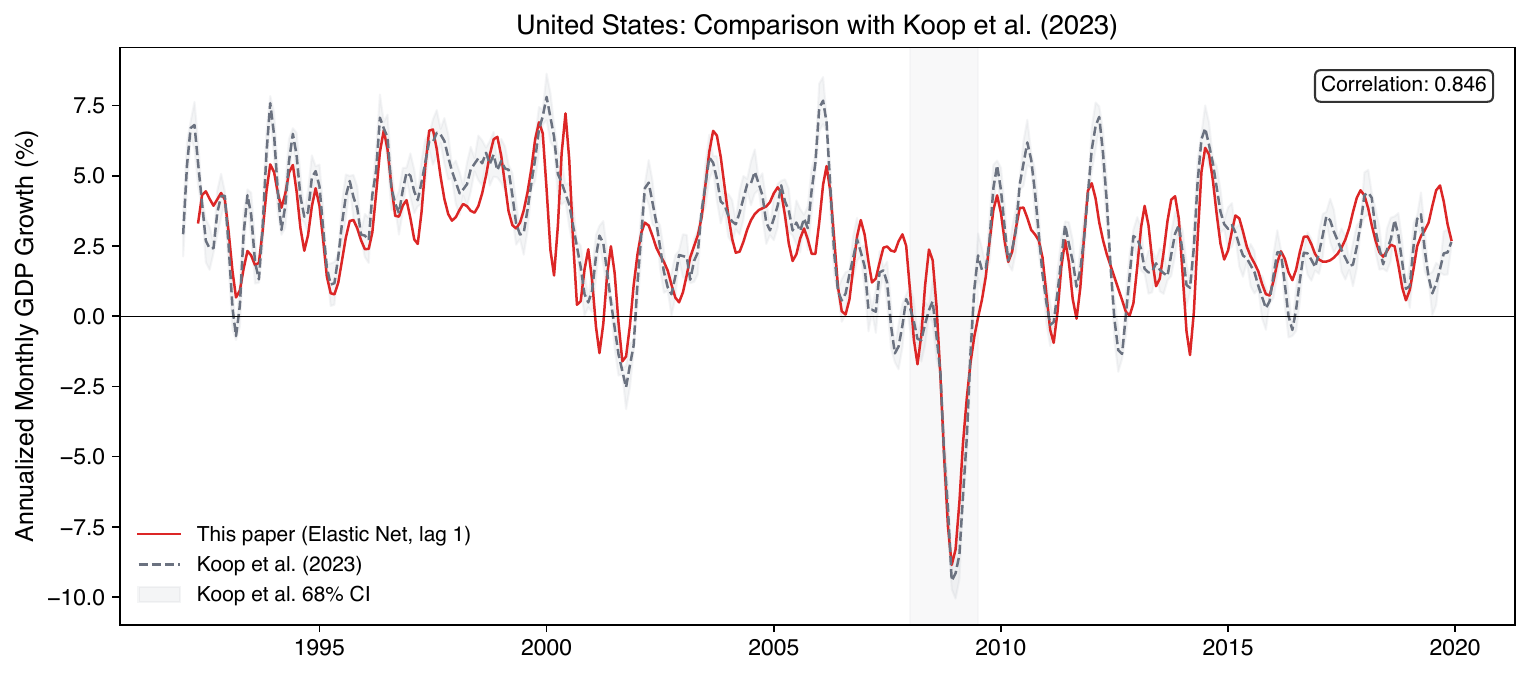}
    \caption{United States: Comparison with \citet{koop2023reconciled}}
    \label{fig:us_benchmark_koop}
    \vspace{0.3em}
    \begin{minipage}{1\textwidth}
    {\footnotesize \textit{Notes}: Annualized monthly GDP growth from Elastic Net (lag~1) versus \citet{koop2023reconciled}. Shaded band: Koop et al.\ 68\% credible interval. Correlation over the overlapping period: 0.85.}
    \end{minipage}
\end{figure}

\Cref{fig:us_benchmark_koop} compares our annualized monthly GDP growth estimates with those of \citet{koop2023reconciled} over the overlapping period. The two series track each other closely, with a correlation of 0.85. Both series capture the major macroeconomic episodes, including the 2001 recession, the GFC, and the subsequent recovery. Our estimates tend to be somewhat smoother, reflecting the use of a smaller set of monthly indicators compared to the Bayesian MF-VAR approach of \citet{koop2023reconciled}. The Koop et al.\ series ends in 2019, so the COVID-19 period cannot be compared.

The Office for National Statistics publishes official monthly GDP estimates from January 1997 onward, indexed to 100 at October 2022. For comparison, we rescale our predicted monthly GDP series to match this base and focus on the overlapping period.

\subsubsection{United Kingdom: Comparison with ONS Monthly GDP}

\begin{figure}[htbp]
    \centering
    \includegraphics[width=\textwidth]{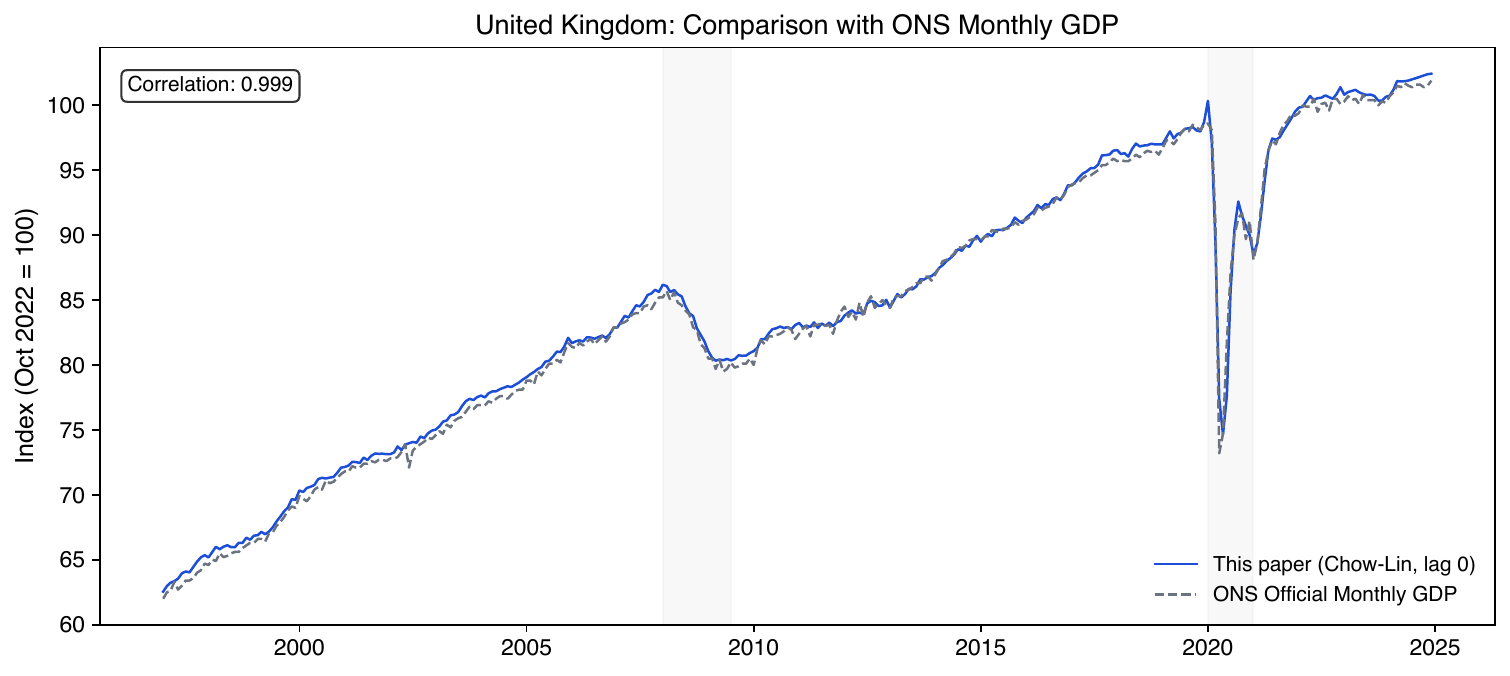}
    \caption{United Kingdom: Comparison with ONS Monthly GDP}
    \label{fig:uk_benchmark_ons}
    \vspace{0.3em}
    \begin{minipage}{1\textwidth}
    {\footnotesize \textit{Notes}: Monthly GDP level from Chow-Lin (lag~0) versus ONS official series. Both indexed to 100 at October 2022. Correlation: 0.999. All four models produce nearly identical monthly series for the UK because quarterly $R^2 < 0.10$; reconciliation dominates.}
    \end{minipage}
\end{figure}

\Cref{fig:uk_benchmark_ons} compares our estimated monthly GDP level with the ONS official series. We use the Chow-Lin estimate for this comparison because all four models achieve similarly low quarterly $R^2$ (0.03--0.09) for the United Kingdom; at this level, the reconciliation step does the majority of the work in distributing quarterly GDP across months, and the choice of regression model has little effect on the final monthly series.\footnote{The best-performing model for the United Kingdom is MLP at lag~0 ($R^2 = 0.09$), but the resulting monthly GDP series is virtually identical to those from the other three models. The correlation between MLP and Chow-Lin monthly estimates exceeds 0.999.} The correlation between our estimates and the ONS series is 0.999, indicating near-perfect agreement in levels. Both series capture the GFC contraction, the subsequent recovery, and the sharp COVID-19 decline. The close agreement confirms that the reconciliation step in \eqref{eq:reconciliation}, which enforces exact consistency with observed quarterly GDP, anchors the monthly estimates to the national accounts even when the model's intra-quarter signal is noisy.

\subsection{What Drives the Gains? \label{sub:drivers}}

The theoretical results in \Cref{sec:theory} identify two potential sources of improvement over the classical \citet{chow1971best} benchmark: nonlinearity (\Cref{prop:bias}) and regularization (\Cref{prop:ridge}). The empirical results allow us to disentangle their contributions.

\paragraph{Regularization, not nonlinearity.} The dominant pattern across countries and lag specifications is that regularization is the primary source of improvement. Elastic Net, a linear model with $\ell_1 + \ell_2$ penalties, achieves the best overall performance (United States, lag~1: $R^2 = 0.87$) and is the only model that consistently improves with the addition of lagged indicators. The nonlinear models (MLP and XGBoost) do not systematically outperform the linear benchmark: the MLP is the best model in only two configurations (Germany lag~1 and United Kingdom lag~0), and in neither case does it significantly outperform the alternatives in Diebold-Mariano tests.

\paragraph{The Chow-Lin curse of dimensionality.} The Chow-Lin method performs well with a small number of contemporaneous indicators (United States lag~0: $R^2 = 0.72$) but degrades dramatically when lagged indicators are added ($R^2 = -1.07$ at lag~2). This pattern is replicated in China, where Chow-Lin is significantly outperformed by all other models at lag~2 in the Diebold-Mariano test. The mechanism is precisely the variance inflation described in \Cref{prop:ridge}: as the number of regressors approaches the sample size, the unpenalized GLS estimator becomes unstable.

\paragraph{Nonlinearity in small samples.} The failure of nonlinear models is consistent with \Cref{rem:finite}. With 60--130 quarterly observations, the MLP and XGBoost cannot reliably estimate their flexible functional forms. The MLP produces negative $R^2$ in most United States lag specifications, indicating that it fits noise rather than signal. The sole exception is Germany at lag~1 ($R^2 = 0.31$), where the MLP captures some dynamic structure that the linear models miss. This suggests that nonlinearity may become valuable as longer time series become available, but is not yet a reliable tool for quarterly GDP disaggregation.

\paragraph{The role of reconciliation.} The benchmark comparisons in \Cref{sub:benchmark} reveal an important feature of the framework: the reconciliation step in \eqref{eq:reconciliation} provides a floor on the quality of monthly estimates regardless of the regression model's predictive accuracy. For the United Kingdom, where all models achieve quarterly $R^2$ below 0.10, the reconciliation constraint $M\hat{y} = z$ forces the monthly estimates to be consistent with observed quarterly GDP, producing a correlation of 0.999 with the ONS official series. In this regime, reconciliation does nearly all the work: the model's intra-quarter signal is weak, and the monthly growth distribution is determined primarily by the quarterly constraints. By contrast, for the United States, where Elastic Net achieves $R^2 = 0.87$, the model contributes a meaningful intra-quarter signal that reconciliation refines rather than overrides. This distinction has a practical implication: model selection matters most when quarterly prediction accuracy is high, because only then does the regression step provide information beyond what reconciliation alone can deliver.

\begin{figure}[htbp]
    \centering
    \includegraphics[width=\textwidth]{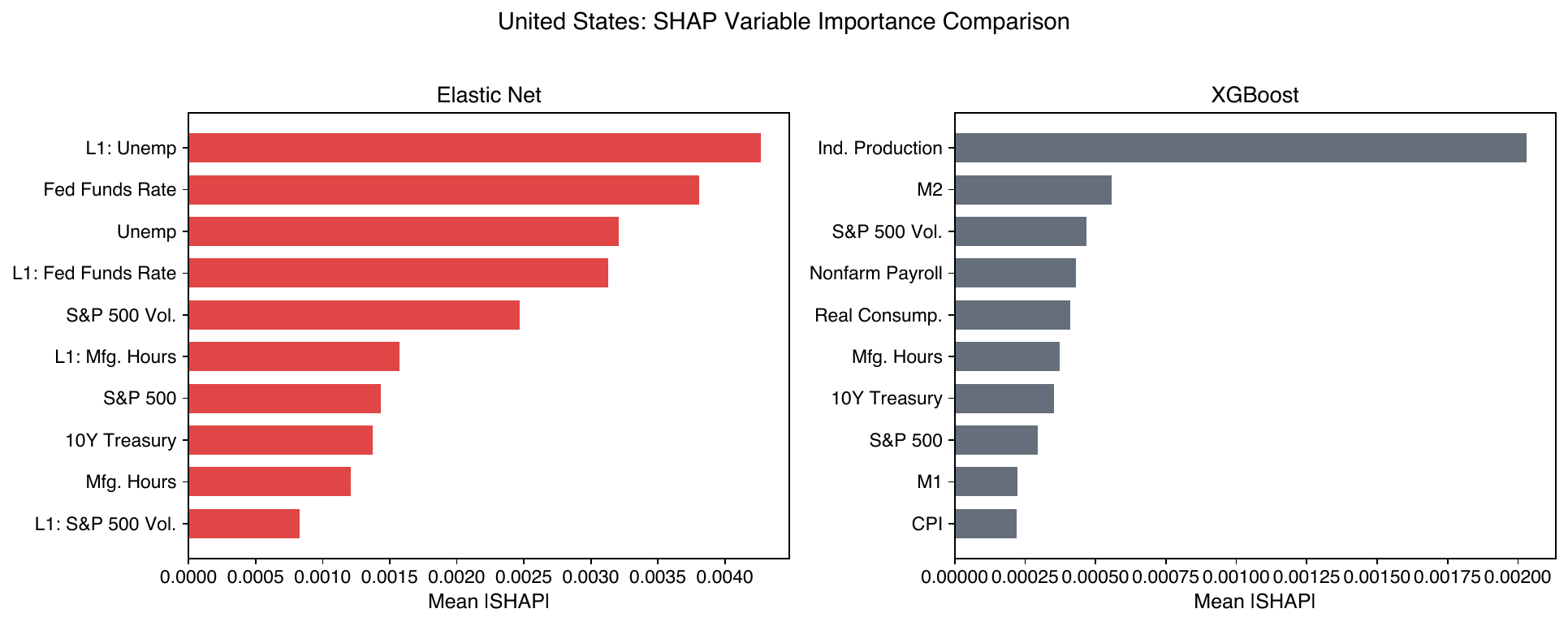}
    \caption{United States: SHAP Variable Importance Comparison}
    \label{fig:shap_comparison}
    \vspace{0.3em}
    \begin{minipage}{1\textwidth}
    {\footnotesize \textit{Notes}: Mean absolute SHAP values for Elastic Net (lag~1, left) and XGBoost (lag~0, right), each at its best-performing lag specification. Top 10 variables shown. Elastic Net emphasizes labor market and financial variables; XGBoost is dominated by industrial production.}
    \end{minipage}
\end{figure}

Second, we use SHAP values to trace which variables contribute most to each model's predictions and how these contributions differ across models. \Cref{fig:shap_comparison} compares the SHAP variable importance rankings for Elastic Net (lag~1) and XGBoost (lag~0) in the United States. We compare each model at its best-performing lag specification to understand what drives each model when it is operating at its strongest.\footnote{The qualitative pattern is similar when both models are compared at lag~0: XGBoost remains dominated by industrial production, while Elastic Net shifts weight toward contemporaneous unemployment and the federal funds rate.} The two models rely on strikingly different information sets. Elastic Net places the greatest weight on labor market indicators (lagged unemployment, the federal funds rate) and financial conditions (S\&P~500 volatility), with lagged variables playing a prominent role. XGBoost, by contrast, is dominated by industrial production, with monetary aggregates (M2) and financial volatility as secondary contributors. Despite these differences in variable weighting, XGBoost does not translate its reliance on real-activity indicators into superior predictive accuracy, consistent with the bias-variance interpretation in \Cref{rem:finite}.

\paragraph{Practical implications.} These findings yield three actionable recommendations for practitioners constructing monthly GDP estimates. First, when a small number of contemporaneous indicators are available, the classical Chow-Lin method remains the appropriate choice; its simplicity is an advantage, not a limitation. Second, when richer information sets are available (multiple indicators with lags), Elastic Net should be preferred over Chow-Lin, as the unpenalized estimator becomes unreliable in high dimensions. Third, nonlinear models such as MLP and XGBoost should not be expected to improve disaggregation accuracy with current quarterly sample sizes; their theoretical advantages require substantially more data to be realized in practice.

\section{Conclusion \label{sec:conclusion}}

This paper proposes a modular framework for temporal disaggregation of GDP in which any supervised learning model can serve as the regression step, with quarterly consistency enforced through Mariano-Murasawa reconciliation. We evaluate four models within this framework across four countries: the classical Chow-Lin method, Elastic Net, XGBoost, and a Multi-Layer Perceptron.

Our theoretical analysis identifies two potential sources of improvement over the classical benchmark. \Cref{prop:bias} shows that linear disaggregation is biased under regime-switching data-generating processes, motivating flexible nonlinear alternatives. \Cref{prop:ridge} shows that regularization reduces mean squared error relative to the unpenalized estimator, motivating penalized linear models. The empirical results reveal that, in the current data environment, the second mechanism dominates: Elastic Net at lag~1 achieves $R^2 = 0.87$ for the United States, substantially outperforming both the Chow-Lin benchmark and the nonlinear alternatives.

The key finding is that the gains from machine learning in temporal disaggregation stem from regularization rather than nonlinearity. The Chow-Lin method, despite its theoretical elegance, is vulnerable to the curse of dimensionality when many indicators and their lags are used simultaneously. Elastic Net addresses this vulnerability through $\ell_1 + \ell_2$ penalization, enabling the use of richer information sets without the variance explosion that afflicts the unpenalized estimator. Nonlinear models, while theoretically capable of capturing regime-dependent dynamics, are unable to realize this advantage with the sample sizes available in quarterly macroeconomic data. A further finding is that the Mariano-Murasawa reconciliation step provides a quality floor: even when the regression model has low predictive power, the quarterly consistency constraint anchors monthly estimates to the national accounts, as demonstrated by the near-perfect correlation (0.999) with the ONS official series for the United Kingdom.

Several avenues for future research emerge. First, as longer time series become available, the bias-variance tradeoff may shift in favor of nonlinear models; revisiting this comparison with additional decades of data would be informative. Second, mixed-frequency approaches that train directly on monthly data, rather than aggregating to quarterly frequency, would increase the effective sample size and may unlock the potential of neural network architectures. Third, the modular framework proposed here can accommodate any future model class, including transformer architectures or other deep learning methods, as they mature for small-sample time series applications.

\clearpage
\bibliographystyle{ecta}
\bibliography{ref}

@article{ghosh2023machine,
  title={{A Machine Learning Approach to {GDP} Nowcasting: An Emerging Market Experience}},
  author={Ghosh, Saurabh and Ranjan, Abhishek},
  journal={Bulletin of Monetary Economics and Banking},
  volume={26},
  pages={33--54},
  year={2023},
  publisher={Bank Indonesia, Central Banking Research Department}
}

@article{mitchell2005indicator,
  title={{An Indicator of Monthly {GDP} and an Early Estimate of Quarterly {GDP} Growth}},
  author={Mitchell, James and Smith, Richard J and Weale, Martin R and Wright, Stephen and Salazar, Eduardo L},
  journal={The Economic Journal},
  volume={115},
  number={501},
  pages={F108--F129},
  year={2005},
  publisher={Oxford University Press Oxford, UK}
}

@article{jansen2016forecasting,
  title={{Forecasting and Nowcasting Real {GDP}: Comparing Statistical Models and Subjective Forecasts}},
  author={Jansen, W Jos and Jin, Xiaowen and de Winter, Jasper M},
  journal={International Journal of Forecasting},
  volume={32},
  number={2},
  pages={411--436},
  year={2016},
  publisher={Elsevier}
}

@article{giannone2008nowcasting,
  title={{Nowcasting: The Real-Time Informational Content of Macroeconomic Data}},
  author={Giannone, Domenico and Reichlin, Lucrezia and Small, David},
  journal={Journal of Monetary Economics},
  volume={55},
  number={4},
  pages={665--676},
  year={2008},
  publisher={Elsevier}
}

@article{koop2023reconciled,
  title={{Reconciled Estimates of Monthly {GDP} in the United States}},
  author={Koop, Gary and McIntyre, Stuart and Mitchell, James and Poon, Aubrey},
  journal={Journal of Business \& Economic Statistics},
  volume={41},
  number={2},
  pages={563--577},
  year={2023},
  publisher={Taylor \& Francis}
}

@article{chu2023comparing,
  title={{Comparing Out-of-Sample Performance of Machine Learning Methods to Forecast {US} {GDP} Growth}},
  author={Chu, Ba and Qureshi, Shafiullah},
  journal={Computational Economics},
  volume={62},
  number={4},
  pages={1567--1609},
  year={2023},
  publisher={Springer}
}

@article{said1984testing,
  title={{Testing for Unit Roots in Autoregressive-Moving Average Models of Unknown Order}},
  author={Said, Said E and Dickey, David A},
  journal={Biometrika},
  volume={71},
  number={3},
  pages={599--607},
  year={1984},
  publisher={Oxford University Press}
}

@book{james2023introduction,
  title={{An Introduction to Statistical Learning with Applications in Python}},
  author={James, Gareth and Witten, Daniela and Hastie, Trevor and Tibshirani, Robert and Taylor, Jonathan},
  year={2023},
  publisher={Springer}
}

@book{hansen2022econometrics,
  title={{Econometrics}},
  author={Hansen, Bruce},
  year={2022},
  publisher={Princeton University Press}
}

@article{lundberg2017unified,
  title={{A Unified Approach to Interpreting Model Predictions}},
  author={Lundberg, Scott M and Lee, Su-In},
  journal={Advances in Neural Information Processing Systems},
  volume={30},
  year={2017}
}

@article{koop2021nowcasting,
  title={{Nowcasting ‘True’ Monthly {US} {GDP} During the Pandemic}},
  author={Koop, Gary and McIntyre, Stuart and Mitchell, James and Poon, Aubrey},
  journal={National Institute Economic Review},
  volume={256},
  pages={44--70},
  year={2021},
  publisher={Cambridge University Press}
}

@article{mariano2010coincident,
  title={{A Coincident Index, Common Factors, and Monthly Real {GDP}}},
  author={Mariano, Roberto S and Murasawa, Yasutomo},
  journal={Oxford Bulletin of Economics and Statistics},
  volume={72},
  number={1},
  pages={27--46},
  year={2010},
  publisher={Wiley Online Library}
}

@article{brave2019new,
  title={{A New “Big Data” Index of {US} Economic Activity}},
  author={Brave, Scott A and Butters, R Andrew and Kelley, David and others},
  journal={Economic Perspectives},
  volume={43},
  number={1},
  pages={1--30},
  year={2019}
}

@article{denton1971adjustment,
  title={{Adjustment of Monthly or Quarterly Series to Annual Totals: An Approach Based on Quadratic Minimization}},
  author={Denton, Frank T},
  journal={Journal of the American Statistical Association},
  volume={66},
  number={333},
  pages={99--102},
  year={1971},
  publisher={Taylor \& Francis}
}

@article{kingma2014adam,
  title={{Adam: A Method for Stochastic Optimization}},
  author={Kingma, Diederik P},
  journal={arXiv preprint arXiv:1412.6980},
  year={2014}
}

@inproceedings{chen2016xgboost,
  title={{XGBoost: A Scalable Tree Boosting System}},
  author={Chen, Tianqi and Guestrin, Carlos},
  booktitle={Proceedings of the 22nd {ACM} {SIGKDD} International Conference on Knowledge Discovery and Data Mining},
  pages={785--794},
  year={2016}
}

@article{rumelhart1986learning,
  title={{Learning Representations by Back-propagating Errors}},
  author={Rumelhart, David E and Hinton, Geoffrey E and Williams, Ronald J},
  journal={Nature},
  volume={323},
  number={6088},
  pages={533--536},
  year={1986},
  publisher={Nature Publishing Group UK London}
}

@article{hochreiter1997long,
  title={{Long Short-term Memory}},
  author={Hochreiter, Sepp and Schmidhuber, J{\"u}rgen},
  journal={Neural Computation},
  volume={9},
  number={8},
  pages={1735--1780},
  year={1997},
  publisher={MIT press}
}

@article{zou2005regularization,
  title={{Regularization and Variable Selection via the Elastic Net}},
  author={Zou, Hui and Hastie, Trevor},
  journal={Journal of the Royal Statistical Society Series B: Statistical Methodology},
  volume={67},
  number={2},
  pages={301--320},
  year={2005},
  publisher={Oxford University Press}
}

@article{mariano2003new,
  title={{A New Coincident Index of Business Cycles Based on Monthly and Quarterly Series}},
  author={Mariano, Roberto S and Murasawa, Yasutomo},
  journal={Journal of Applied Econometrics},
  volume={18},
  number={4},
  pages={427--443},
  year={2003},
  publisher={Wiley Online Library}
}

@article{chow1971best,
  title={{Best Linear Unbiased Interpolation, Distribution, and Extrapolation of Time Series by Related Series}},
  author={Chow, Gregory C and Lin, {An-loh}},
  journal={The Review of Economics and Statistics},
  pages={372--375},
  year={1971},
  publisher={JSTOR}
}

@article{litterman1983random,
  title={{A Random Walk, Markov Model for the Distribution of Time Series}},
  author={Litterman, Robert B},
  journal={Journal of Business \& Economic Statistics},
  volume={1},
  number={2},
  pages={169--173},
  year={1983},
  publisher={Taylor \& Francis}
}

@article{rawski2001happening,
  title={{What is Happening to China's GDP Statistics?}},
  author={Rawski, Thomas G},
  journal={China Economic Review},
  volume={12},
  number={4},
  pages={347--354},
  year={2001},
  publisher={Elsevier}
}

@article{holz2014quality,
  title={{The Quality of China's GDP Statistics}},
  author={Holz, Carsten A},
  journal={China Economic Review},
  volume={30},
  pages={309--338},
  year={2014},
  publisher={Elsevier}
}

@article{scruton2018introducing,
  title={{Introducing a New Publication Model for GDP}},
  author={Scruton, James and O’Donnell, Molly and Dey-Chowdhury, Sumit},
  journal={Office for National Statistics},
  pages={1--14},
  year={2018}
}

@article{goulet2022machine,
  title={{How is Machine Learning Useful for Macroeconomic Forecasting?}},
  author={Goulet Coulombe, Philippe and Leroux, Maxime and Stevanovic, Dalibor and Surprenant, St{\'e}phane},
  journal={Journal of Applied Econometrics},
  volume={37},
  number={5},
  pages={920--964},
  year={2022},
  publisher={Wiley Online Library}
}

@article{richardson2021nowcasting,
  title={{Nowcasting GDP using Machine-learning Algorithms: A Real-time Assessment}},
  author={Richardson, Adam and van Florenstein Mulder, Thomas and Vehbi, Tu{\u{g}}rul},
  journal={International Journal of Forecasting},
  volume={37},
  number={2},
  pages={941--948},
  year={2021},
  publisher={Elsevier}
}

@article{diebold1995comparing,
  title={Comparing Predictive Accuracy},
  author={Diebold, Francis X and Mariano, Roberto S},
  journal={Journal of Business \& Economic Statistics},
  volume={13},
  number={3},
  pages={253--263},
  year={1995}
}

@article{chan2023high,
  title={High-Dimensional Conditionally {G}aussian State Space Models with Missing Data},
  author={Chan, Joshua CC and Poon, Aubrey and Zhu, Dan},
  journal={Journal of Econometrics},
  volume={236},
  number={1},
  pages={105468},
  year={2023}
}

@article{cleveland1990stl,
  title={{STL}: A Seasonal-Trend Decomposition Procedure Based on {L}oess},
  author={Cleveland, Robert B and Cleveland, William S and McRae, Jean E and Terpenning, Irma},
  journal={Journal of Official Statistics},
  volume={6},
  number={1},
  pages={3--73},
  year={1990}
}

@article{fernandez1981methodological,
  title={A Methodological Note on the Estimation of Time Series},
  author={Fern{\'a}ndez, Roque B},
  journal={Review of Economics and Statistics},
  volume={63},
  number={3},
  pages={471--476},
  year={1981}
}

@article{hornik1989multilayer,
  title={{Multilayer Feedforward Networks are Universal Approximators}},
  author={Hornik, Kurt and Stinchcombe, Maxwell and White, Halbert},
  journal={Neural Networks},
  volume={2},
  number={5},
  pages={359--366},
  year={1989},
  publisher={Elsevier}
}

@article{hansen2000sample,
  title={{Sample Splitting and Threshold Estimation}},
  author={Hansen, Bruce E},
  journal={Econometrica},
  volume={68},
  number={3},
  pages={575--603},
  year={2000},
  publisher={Wiley Online Library}
}

@article{fernald2016reassessing,
  title={Is {China} Faking its {GDP} Figures? {Evidence} from Trading Partner Data},
  author={Fernald, John and Hsu, Eric and Spiegel, Mark M.},
  journal={Federal Reserve Bank of San Francisco Working Paper},
  year={2019}
}

@article{hoerl1970ridge,
  title={Ridge Regression: Biased Estimation for Nonorthogonal Problems},
  author={Hoerl, Arthur E. and Kennard, Robert W.},
  journal={Technometrics},
  volume={12},
  number={1},
  pages={55--67},
  year={1970}
}

\clearpage
\appendix
\pagestyle{fancy}
\fancyhf{}
\fancyhead[LE,LO]{Online Appendix: For Online Publication Only}
\fancyhead[RE,RO]{\thepage}
\setcounter{footnote}{0}
\setlength{\headsep}{0.35in}

\section{Proofs \label{app:proofs}}
\renewcommand\theequation{A\arabic{equation}} \setcounter{equation}{0}

\subsection{Proof of \Cref{prop:bias}}

Under the DGP in (8), the unconditional expectation of $Y_q$ given $X_q$ is
\begin{flalign}
    \mathbb{E}[Y_q \mid X_q] = (1 - \pi)\,\beta_1^\top X_q + \pi\,\beta_2^\top X_q = \bar{\beta}^\top X_q,
\end{flalign}
where $\bar{\beta} = (1 - \pi)\,\beta_1 + \pi\,\beta_2$. Since OLS minimizes the unconditional squared prediction error, $\hat{\beta}_{\text{OLS}} \xrightarrow{p} \bar{\beta}$, establishing (i). For (ii), in the crisis regime the true conditional expectation is $\beta_2^\top X_q$, so the prediction error is
\begin{flalign}
    \beta_2^\top X_q - \bar{\beta}^\top X_q = (\beta_2 - \bar{\beta})^\top X_q = (1 - \pi)(\beta_2 - \beta_1)^\top X_q.
\end{flalign}
Taking expectations conditional on $s_q = 1$ yields (ii). Part (iii) follows by taking $\pi \to 0$. \hfill $\square$

\subsection{Proof of \Cref{prop:ridge}}

The MSE of the ridge estimator decomposes as
\begin{flalign}
    \text{MSE}(\hat{\beta}_\lambda) = \sigma^2 \sum_{j=1}^k \frac{d_j}{(d_j + \lambda)^2} + \lambda^2 \sum_{j=1}^k \frac{\alpha_j^2}{(d_j + \lambda)^2},
\end{flalign}
where $d_1, \ldots, d_k$ are the eigenvalues of $X^\top X$ and $\alpha_j = v_j^\top \beta$ with $v_j$ the corresponding eigenvectors. The first term is the variance (strictly decreasing in $\lambda$), the second is the squared bias (strictly increasing in $\lambda$ from zero). Since the derivative of $\text{MSE}(\hat{\beta}_\lambda)$ with respect to $\lambda$ at $\lambda = 0$ is $-2\sigma^2 \sum_j d_j^{-2} < 0$, there exists $\lambda^* > 0$ with strictly lower MSE than OLS. This is the result of \citet{hoerl1970ridge}. \hfill $\square$

\section{Data Descriptions \label{app:data}}
\renewcommand{\thetable}{B\arabic{table}} \setcounter{table}{0}
\renewcommand{\thefigure}{B\arabic{figure}} \setcounter{figure}{0}
\renewcommand\theequation{B\arabic{equation}} \setcounter{equation}{0}

All data used in this study are publicly available from open-source platforms. Most datasets are retrieved via the FRED API and the Yahoo Finance API. The processed datasets used in this study are stored as \texttt{master\_<country>.csv} files in the replication package. All summary statistics are based on raw data prior to any transformation.

\begin{longtable}{llcc}
\caption{Dataset Description by Country} \label{tab:dataset_all} \\
\toprule
Variable & Code & Unit Root & Transform. \\
\midrule
\endfirsthead
\toprule
Variable & Code & Unit Root & Transform. \\
\midrule
\endhead
\bottomrule
\endfoot
\textit{China} & & & \\
CPI (All Items, Growth) & \texttt{CPALTT01CNM657N} & No & $(X_m - \mu_j)/\sigma_j$ \\
PPI (Furniture, Household) & \texttt{WPU1261} & Yes & $\Delta\log(X_m)$ \\
Total Reserves (excl.\ gold) & \texttt{TRESEGCNM052N} & Yes & $\Delta\log(X_m)$ \\
Exports & \texttt{XTEXVA01CNM664S} & Yes & $\Delta\log(X_m)$ \\
Imports & \texttt{XTIMVA01CNM664S} & Yes & $\Delta\log(X_m)$ \\
Exchange Rate (RMB/USD) & \texttt{EXCHUS} & Yes & $\Delta\log(X_m)$ \\
Policy Uncertainty & \texttt{CHNMAINLANDEPU} & Yes & $\Delta\log(X_m)$ \\
US Imports from China & \texttt{IMPCH} & Yes & $\Delta\log(X_m)$ \\
Effective Exchange Rate & \texttt{RBCNBIS} & Yes & $\Delta\log(X_m)$ \\
SSEC Index & \texttt{Investing.com} & Yes & $\Delta\log(X_m)$ \\
Real GDP & \texttt{CHNGDPNQDSMEI} & --- & $\Delta\log(\text{GDP}_q)$ \\
\textit{Germany} & & & \\
CPI (All Items) & \texttt{DEUCPIALLMINMEI} & Yes & $\Delta\log(X_m)$ \\
Unemployment Rate & \texttt{LRHUTTTTDEM156S} & Yes & $\Delta\log(X_m)$ \\
Production Volume & \texttt{DEUPROINDMISMEI} & Yes & $\Delta\log(X_m)$ \\
Retail Trade & \texttt{DEUSARTMISMEI} & Yes & $\Delta\log(X_m)$ \\
Exports & \texttt{XTEXVA01DEM664S} & Yes & $\Delta\log(X_m)$ \\
Imports & \texttt{XTIMVA01DEM664S} & Yes & $\Delta\log(X_m)$ \\
Total Reserves (excl.\ gold) & \texttt{TRESEGDEM052N} & Yes & $\Delta\log(X_m)$ \\
Exchange Rate & \texttt{CCUSMA02DEM650N} & No & $(X_m - \mu_j)/\sigma_j$ \\
Effective Exchange Rate & \texttt{RNDEBIS} & Yes & $\Delta\log(X_m)$ \\
Policy Uncertainty & \texttt{EUEPUINDXM} & Yes & $\Delta\log(X_m)$ \\
DAX Index & \texttt{Yahoo Finance} & Yes & $\Delta\log(X_m)$ \\
Price Competitiveness & \texttt{Bundesbank} & Yes & $\Delta\log(X_m)$ \\
Real GDP & \texttt{CLVMNACSCAB1GQDE} & --- & $\Delta\log(\text{GDP}_q)$ \\
\textit{United Kingdom} & & & \\
Share Prices & \texttt{SPASTT01GBM661N} & Yes & $\Delta\log(X_m)$ \\
CPI (All Items) & \texttt{GBRCPIALLMINMEI} & Yes & $\Delta\log(X_m)$ \\
Production Volume & \texttt{GBRPROINDMISMEI} & Yes & $\Delta\log(X_m)$ \\
Interest Rate (Gov Bond) & \texttt{INTGSBGBM193N} & Yes & $\Delta X_m$ \\
Interest Rate (10Y Gov) & \texttt{IRLTLT01GBM156N} & Yes & $\Delta X_m$ \\
Total Reserves (excl.\ gold) & \texttt{TRESEGGBM052N} & Yes & $\Delta\log(X_m)$ \\
Exchange Rate (GBP/USD) & \texttt{EXUSUK} & Yes & $\Delta\log(X_m)$ \\
Imports & \texttt{XTIMVA01GBM664S} & Yes & $\Delta\log(X_m)$ \\
Exports & \texttt{XTEXVA01GBM664S} & Yes & $\Delta\log(X_m)$ \\
Unemployment Rate & \texttt{ONS} & Yes & $\Delta\log(X_m)$ \\
FTSE 100 Index & \texttt{Yahoo Finance} & Yes & $\Delta\log(X_m)$ \\
M1 Money Stock & \texttt{BOE} & Yes & $\Delta\log(X_m)$ \\
Employment Rate & \texttt{ONS} & Yes & $\Delta\log(X_m)$ \\
Real GDP & \texttt{NGDPRSAXDCGBQ} & --- & $\Delta\log(\text{GDP}_q)$ \\
\textit{United States} & & & \\
Avg.\ Weekly Hours, Manuf. & \texttt{AWHMAN} & No & $(X_m - \mu_j)/\sigma_j$ \\
CPI (All Items) & \texttt{CPIAUCSL} & Yes & $\Delta\log(X_m)$ \\
Industrial Production & \texttt{INDPRO} & Yes & $\Delta\log(X_m)$ \\
Real Personal Consum. & \texttt{DPCERA3M086SBEA} & Yes & $\Delta\log(X_m)$ \\
Effective Fed Funds Rate & \texttt{FEDFUNDS} & No & $(X_m - \mu_j)/\sigma_j$ \\
Interest Rate (10Y Trea.) & \texttt{GS10} & Yes & $\Delta X_m$ \\
S\&P 500 Volatility & \texttt{VIXCLS} & No & $(X_m - \mu_j)/\sigma_j$ \\
Unemployment Rate & \texttt{UNRATE} & No & $(X_m - \mu_j)/\sigma_j$ \\
Nonfarm Payroll Emp. & \texttt{PAYEMS} & Yes & $\Delta\log(X_m)$ \\
Labor Force Participation & \texttt{CIVPART} & Yes & $\Delta\log(X_m)$ \\
M1 Money Stock & \texttt{M1SL} & Yes & $\Delta\log(X_m)$ \\
M2 Money Stock & \texttt{M2SL} & Yes & $\Delta\log(X_m)$ \\
Net Exports & \texttt{BOPTEXP} & Yes & $\Delta\log(X_m)$ \\
Moody's AAA Bond Yield & \texttt{AAA} & Yes & $\Delta X_m$ \\
S\&P 500 Index & \texttt{Yahoo Finance} & Yes & $\Delta\log(X_m)$ \\
Real GDP & \texttt{GDPC1} & --- & $\Delta\log(\text{GDP}_q)$ \\
\addlinespace
\end{longtable}
\vspace{0.5em}
\noindent{\footnotesize \textit{Notes}: Most series are from FRED. Stock indices are from Yahoo Finance. SSEC Index is from Investing.com. UK M1 and employment data are from BOE/ONS. Germany price competitiveness is from the Deutsche Bundesbank. ``Unit Root'' indicates non-stationarity based on the Augmented Dickey-Fuller test at the 5\% level and economic considerations. $\Delta\log(X_m)$: log-difference; $\Delta X_m$: first-difference (applied to interest rates); $(X_m - \mu_j)/\sigma_j$: Z-score standardization.}

\begin{longtable}{lrrrrr}
\caption{Summary Statistics by Country} \label{tab:data_stat_all} \\
\toprule
Variable & Obs. & Mean & Std.\ Dev. & Min & Max \\
\midrule
\endfirsthead
\toprule
Variable & Obs. & Mean & Std.\ Dev. & Min & Max \\
\midrule
\endhead
\bottomrule
\endfoot
\textit{China} & & & & & \\
CPI (All Items, Growth) & 357 & 0.2483 & 0.7958 & -1.403 & 4.048 \\
PPI (Furniture, Household) & 357 & 220.26 & 43.09 & 156.20 & 330.02 \\
Total Reserves (excl.\ gold) & 357 & 1,808,896 & 1,439,990 & 26,355 & 4,010,834 \\
Exports & 357 & 8.06e+11 & 5.90e+11 & 5.77e+10 & 2.22e+12 \\
Imports & 357 & 6.64e+11 & 4.63e+11 & 6.24e+10 & 1.62e+12 \\
Exchange Rate (RMB/USD) & 357 & 7.382 & 0.8578 & 6.050 & 8.725 \\
Policy Uncertainty & 357 & 145.33 & 112.20 & 10.11 & 661.83 \\
US Imports from China & 357 & 1.51e+11 & 5.90e+10 & 5.00e+10 & 2.89e+11 \\
Effective Exchange Rate & 357 & 82.61 & 13.72 & 52.78 & 106.40 \\
SSEC Index & 357 & 2,256 & 1,017 & 333.92 & 5,955 \\
Real GDP & 119 & 1.14e+13 & 9.59e+12 & 9.38e+11 & 3.36e+13 \\
\textit{Germany} & & & & & \\
CPI (All Items) & 399 & 90.41 & 13.99 & 64.00 & 125.04 \\
Unemployment Rate & 399 & 6.642 & 2.488 & 2.900 & 11.20 \\
Production Volume & 399 & 87.88 & 11.64 & 65.25 & 107.38 \\
Retail Trade & 399 & 100.21 & 7.194 & 90.71 & 122.61 \\
Exports & 399 & 7.25e+10 & 3.16e+10 & 2.58e+10 & 1.39e+11 \\
Imports & 399 & 6.11e+10 & 2.68e+10 & 2.33e+10 & 1.34e+11 \\
Total Reserves (excl.\ gold) & 399 & 65,176 & 15,603 & 39,972 & 120,255 \\
Exchange Rate & 399 & 0.8499 & 0.1077 & 0.6346 & 1.172 \\
Effective Exchange Rate & 399 & 102.61 & 4.319 & 93.90 & 116.96 \\
Policy Uncertainty & 399 & 153.95 & 78.08 & 47.69 & 433.28 \\
DAX Index & 399 & 7,102 & 4,206 & 1,374 & 18,001 \\
Price Competitiveness & 399 & 98.72 & 3.882 & 90.90 & 111.90 \\
Real GDP & 140 & 651,251 & 83,011 & 514,471 & 774,367 \\
\textit{United Kingdom} & & & & & \\
Share Prices & 399 & 83.34 & 22.75 & 31.94 & 120.02 \\
CPI (All Items) & 399 & 87.01 & 18.22 & 57.00 & 131.60 \\
Production Volume & 399 & 96.73 & 9.616 & 75.20 & 113.32 \\
Interest Rate (Gov Bond) & 399 & 4.344 & 2.550 & 0.2094 & 10.59 \\
Interest Rate (10Y Gov) & 399 & 4.342 & 2.547 & 0.2094 & 10.59 \\
Total Reserves (excl.\ gold) & 399 & 82,244 & 46,308 & 28,913 & 176,024 \\
Exchange Rate (GBP/USD) & 399 & 1.558 & 0.2017 & 1.132 & 2.070 \\
Imports & 399 & 2.66e+10 & 1.14e+10 & 9.34e+09 & 5.76e+10 \\
Exports & 399 & 1.96e+10 & 6.43e+09 & 8.30e+09 & 3.44e+10 \\
Unemployment Rate & 399 & 6.228 & 1.867 & 3.600 & 10.70 \\
FTSE 100 Index & 399 & 5,496 & 1,500 & 2,107 & 7,915 \\
M1 Money Stock & 399 & 992,470 & 680,818 & 174,324 & 2,575,027 \\
Employment Rate & 399 & 72.36 & 2.054 & 68.30 & 76.50 \\
Real GDP & 140 & 548,706 & 99,764 & 372,751 & 705,571 \\
\textit{United States} & & & & & \\
Avg.\ Weekly Hours, Manuf. & 407 & 41.19 & 0.5984 & 38.40 & 42.40 \\
CPI (All Items) & 407 & 214.51 & 49.44 & 138.30 & 326.03 \\
Industrial Production & 407 & 91.98 & 11.20 & 61.46 & 104.10 \\
Real Personal Consum. & 407 & 84.24 & 20.58 & 49.07 & 125.53 \\
Effective Fed Funds Rate & 407 & 2.640 & 2.171 & 0.0500 & 6.540 \\
Interest Rate (10Y Trea.) & 407 & 4.018 & 1.731 & 0.6200 & 7.960 \\
S\&P 500 Volatility & 407 & 19.39 & 7.562 & 10.13 & 62.67 \\
Unemployment Rate & 407 & 5.644 & 1.779 & 3.400 & 14.80 \\
Nonfarm Payroll Emp. & 407 & 135,257 & 12,583 & 108,312 & 158,548 \\
Labor Force Participation & 407 & 64.79 & 1.903 & 60.10 & 67.30 \\
M1 Money Stock & 407 & 4,639 & 6,375 & 910.40 & 20,750 \\
M2 Money Stock & 407 & 9,867 & 5,948 & 3,386 & 22,505 \\
Net Exports & 407 & -40,782 & 22,341 & -135,856 & -831.00 \\
Moody's AAA Bond Yield & 407 & 5.402 & 1.583 & 2.140 & 8.680 \\
S\&P 500 Index & 407 & 1,895 & 1,421 & 407.36 & 6,853 \\
Real GDP & 136 & 16,747 & 3,750 & 10,236 & 24,066 \\
\addlinespace
\end{longtable}
\noindent{\footnotesize \textit{Notes}: Summary statistics are calculated using raw values before any transformation (e.g., differencing or standardization).}

\section{Model Specifications \label{app:models}}
\renewcommand{\thetable}{C\arabic{table}} \setcounter{table}{0}
\renewcommand{\thefigure}{C\arabic{figure}} \setcounter{figure}{0}
\renewcommand\theequation{C\arabic{equation}} \setcounter{equation}{0}

This appendix provides detailed specifications of each model used in the paper. All models share the general regression framework $Y_q = f(X_q; \theta) + e_q$ described in \Cref{sub:framework}. We describe each model's functional form, estimation procedure, and hyperparameter tuning strategy.

\subsection{Chow-Lin}

The \citet{chow1971best} method estimates the linear relationship
\begin{flalign}
    Y_q = X_q^\top \beta + u_q,
\end{flalign}
where $\beta \in \mathbb{R}^k$ is the coefficient vector and $u_q$ is the error term. At the monthly frequency, the residuals are assumed to follow a first-order autoregressive process:
\begin{flalign}
    u_m = \rho \, u_{m-1} + \varepsilon_m, \quad \varepsilon_m \sim \text{i.i.d.}(0, \sigma^2_\varepsilon),
\end{flalign}
where $|\rho| < 1$. This implies a monthly covariance matrix of the form
\begin{flalign}
    \Sigma(\rho) = \frac{\sigma^2_\varepsilon}{1 - \rho^2}
    \begin{pmatrix}
    1 & \rho & \rho^2 & \cdots & \rho^{T-1} \\
    \rho & 1 & \rho & \cdots & \rho^{T-2} \\
    \vdots & \vdots & \vdots & \ddots & \vdots \\
    \rho^{T-1} & \rho^{T-2} & \rho^{T-3} & \cdots & 1
    \end{pmatrix},
\end{flalign}
where $T$ is the total number of months in the sample. Given the temporal aggregation matrix $J$ (which sums monthly observations into quarterly totals), the quarterly covariance matrix is $\Omega(\rho) = J \Sigma(\rho) J^\top$. The generalized least squares estimator is then
\begin{flalign}
    \hat{\beta}(\rho) = \left( X_q^\top \Omega(\rho)^{-1} X_q \right)^{-1} X_q^\top \Omega(\rho)^{-1} Y_q.
\end{flalign}
The optimal $\hat{\rho}$ is selected by maximizing the generalized least squares log-likelihood:
\begin{flalign}
    \ell(\rho) = -\frac{Q}{2} \log(2\pi) - \frac{1}{2} \log |\Omega(\rho)| - \frac{1}{2} \hat{u}_q(\rho)^\top \Omega(\rho)^{-1} \hat{u}_q(\rho),
\end{flalign}
where $Q$ is the number of quarters and $\hat{u}_q(\rho) = Y_q - X_q^\top \hat{\beta}(\rho)$. In practice, we search over $\rho \in (-1, 1)$ using numerical optimization. Given $\hat{\beta}$ and $\hat{\rho}$, the disaggregated monthly estimates are obtained by distributing the quarterly residuals across months according to the AR(1) covariance structure.

\subsection{Multi-Layer Perceptron (MLP)}

The architecture and training procedure of the MLP are described in \Cref{sub:models}. The default configuration uses up to 2 hidden layers and 128 neurons per layer, with a maximum of 1,000 training epochs. The candidate activation functions include the Rectified Linear Unit (ReLU), hyperbolic tangent (Tanh), Exponential Linear Unit (ELU), Scaled Exponential Linear Unit (SELU), and Swish. The final layer uses a linear activation function to reflect the continuous nature of GDP growth. Early stopping with a patience of 50 epochs restores the weights from the epoch with the lowest validation loss. In the expanding-window evaluation, the architecture is selected via Bayesian optimization (up to 100 trials) on the initial training window and held fixed for subsequent windows. The search space is deliberately constrained to shallow, narrow architectures to mitigate overfitting given the small effective sample size (60--130 quarterly observations).

\subsection{Elastic Net}

The Elastic Net is a regularized linear regression model that combines the $\ell_1$ penalty of the LASSO and the $\ell_2$ penalty of Ridge regression \citep{zou2005regularization}:
\begin{flalign}
\mathcal{L}(\beta) = \sum_q \left( Y_q - X_q^\top \beta \right)^2 
+ \lambda_1 \sum_{j=1}^k |\beta_j| 
+ \lambda_2 \sum_{j=1}^k \beta_j^2.
\end{flalign}
The regularization parameters are reparameterized as $\lambda_1 = \alpha \rho$ and $\lambda_2 = \alpha(1 - \rho)$, where $\alpha > 0$ controls the overall regularization strength and $\rho \in [0,1]$ determines the mixing ratio. We search over a grid of $\rho$ values $\{0.1, 0.3, 0.5, 0.7, 0.9, 0.95, 0.99, 1.0\}$ and 100 values of $\alpha$ determined by the regularization path, selecting the pair $(\hat{\alpha}, \hat{\rho})$ that minimizes the cross-validated mean squared error.

The Elastic Net can be viewed as a high-dimensional extension of the \citet{chow1971best} framework: both are linear in the relationship between indicators and GDP, but the Elastic Net handles larger sets of potentially correlated regressors through regularization. We employ a nonparametric bootstrap procedure ($B = 5{,}000$ replications) to improve the stability of coefficient estimates for the final monthly GDP disaggregation \citep{hansen2022econometrics, james2023introduction}. During the expanding-window evaluation, bootstrap is not performed; only the regularization parameters from the initial window are reused to reduce computation.

\subsection{XGBoost}

XGBoost (Extreme Gradient Boosting) is a high-performance implementation of the gradient boosting framework that combines an ensemble of decision trees \citep{chen2016xgboost}. The ensemble prediction after $T$ iterations is $\hat{Y}_q = \sum_{t=1}^{T} g_t(X_q)$, where each tree $g_t$ is chosen to minimize a regularized objective. XGBoost incorporates both $\ell_2$ regularization on leaf weights and a penalty on tree complexity to prevent overfitting. Key hyperparameters are optimized via 5-fold cross-validated grid search over 432 combinations.

\section{Algorithms \label{app:algorithms}}

\renewcommand{\thetable}{D\arabic{table}}\setcounter{table}{0}
\renewcommand{\thefigure}{D\arabic{figure}}\setcounter{figure}{0}
\renewcommand\theequation{D\arabic{equation}}\setcounter{equation}{0}

This section illustrates the algorithms described in \Cref{sub:framework}. All algorithms are implemented in Python; the scripts are available in the replication package.\footnote{\href{https://github.com/Yonggeun-Jung/monthly_gdp}{https://github.com/Yonggeun-Jung/monthly\_gdp}}

\subsection{Data Preprocessing}

\begin{algorithm}[H]
\SetAlgoLined
\KwData{Master file \texttt{master\_country.csv} with columns: \texttt{DATE}, $X_1, \dots, X_k$, $Y$; pre-defined \texttt{log\_diff\_cols}, \texttt{diff\_cols}}
\KwResult{Transformed monthly explanatory variables \texttt{X\_prime\_m\_df}}

Load \texttt{master\_country.csv} into DataFrame \texttt{df}\;
Convert \texttt{DATE} column to datetime format\;
Create \texttt{quarter} column by extracting quarter from \texttt{DATE}\;

\tcp{Optional: STL seasonal adjustment (China only)}
\If{\texttt{seasonal\_adjust} is specified}{
    \For{each column $X_j$ to be adjusted}{
        Apply STL decomposition with period $= 12$\;
        Replace $X_j$ with trend $+$ residual components\;
    }
}

Initialize \texttt{X\_prime\_m\_df} with columns \texttt{DATE} and \texttt{quarter}\;

\For{each explanatory variable $X_j \in \{X_1, \dots, X_k\}$}{
    Conduct Augmented Dickey-Fuller test on $X_j$\;
    \uIf{$X_j \in$ \texttt{log\_diff\_cols}}{
        $X'_{m,j} \leftarrow \log(X_{m,j}) - \log(X_{m-1,j})$ \tcp*{Log-difference}
    }
    \uElseIf{$X_j \in$ \texttt{diff\_cols}}{
        $X'_{m,j} \leftarrow X_{m,j} - X_{m-1,j}$ \tcp*{First-difference (e.g., interest rates)}
    }
    \Else{
        $X'_{m,j} \leftarrow X_{m,j}$ \tcp*{Levels (stationary variables)}
    }
    Add $X'_{m,j}$ to \texttt{X\_prime\_m\_df}\;
}
\caption{Monthly Transformation of Explanatory Variables}
\label{algo1}
\end{algorithm}

In Algorithm~\ref{algo1}, we preprocess the dataset. The primary task is to determine how to transform each explanatory variable $X_j$. We conduct the Augmented Dickey-Fuller test \citep{said1984testing} to detect unit roots, but the final classification is guided by economic considerations. For China, we apply STL \citep{cleveland1990stl} to all explanatory variables before any transformation, to address residual seasonality.

\begin{algorithm}[H]
\SetAlgoLined
\KwData{Transformed monthly \texttt{X\_prime\_m\_df}; \texttt{log\_diff\_cols}; \texttt{diff\_cols}}
\KwResult{Quarterly dataset \texttt{X\_q\_processed}, \texttt{Y\_q\_processed}}

\texttt{quarters} $\leftarrow$ unique sorted quarters from \texttt{X\_prime\_m\_df}\;

\For{each quarter $q$ in \texttt{quarters}}{
    \For{each variable $j$ in explanatory variables}{
        \uIf{$j \in$ \texttt{log\_diff\_cols} $\cup$ \texttt{diff\_cols}}{
            $X_{q,j} \leftarrow \sum_{m \in q} X'_{m,j}$ \tcp*{Cumulative change}
        }
        \Else{
            $X_{q,j} \leftarrow \frac{1}{3} \sum_{m \in q} X'_{m,j}$ \tcp*{Quarterly average}
        }
    }
}

$Y_q \leftarrow \log(\text{GDP}_q) - \log(\text{GDP}_{q-1})$ \tcp*{Quarterly GDP growth}

\caption{Quarterly Aggregation and Final Processing}
\label{algo2}
\end{algorithm}

\subsection{Expanding-Window Evaluation}

All models are evaluated using the expanding-window protocol described in \Cref{sub:evaluation}. The following algorithm applies to each model.

\begin{algorithm}[H]
\SetAlgoLined
\KwData{\texttt{X\_q\_processed}, \texttt{Y\_q\_processed}, model class $\mathcal{M}$, initial ratio $r_0 = 0.5$}
\KwResult{Out-of-sample predictions $\{\hat{Y}_t\}_{t=t_0}^{N}$}

$N \leftarrow$ number of quarters\;
$t_0 \leftarrow \lceil r_0 \cdot N \rceil$ \tcp*{Initial training window size}
$h^* \leftarrow \texttt{NULL}$ \tcp*{Best hyperparameters (to be fixed after first window)}

\For{$t = t_0$ to $N-1$}{
    $\mathcal{D}_{\text{train}} \leftarrow \{(X_q, Y_q)\}_{q=1}^{t}$ \tcp*{Expanding training window}
    $(X_{\text{test}}, Y_{\text{test}}) \leftarrow (X_{t+1}, Y_{t+1})$ \tcp*{One-step-ahead target}
    
    \tcp{Scale level variables (fit on training window only)}
    Fit \texttt{StandardScaler} on non-differenced columns of $\mathcal{D}_{\text{train}}$\;
    Apply scaler to both $\mathcal{D}_{\text{train}}$ and $X_{\text{test}}$\;
    
    \tcp{For neural networks: internal validation split}
    $\mathcal{D}_{\text{val}} \leftarrow$ last 20\% of $\mathcal{D}_{\text{train}}$\;
    $\mathcal{D}_{\text{tr}} \leftarrow$ first 80\% of $\mathcal{D}_{\text{train}}$\;
    
    \tcp{Architecture selection vs.\ fixed retraining}
    \eIf{$t = t_0$}{
        $\hat{f}_t, \; h^* \leftarrow \mathcal{M}.\text{fit}(\mathcal{D}_{\text{tr}}, \mathcal{D}_{\text{val}})$ \tcp*{Full hyperparameter search}
    }{
        $\hat{f}_t \leftarrow \mathcal{M}.\text{fit\_fixed}(\mathcal{D}_{\text{tr}}, \mathcal{D}_{\text{val}}, h^*)$ \tcp*{Re-estimate weights only}
    }
    $\hat{Y}_{t+1} \leftarrow \hat{f}_t(X_{\text{test}})$\;
}

\caption{Expanding-Window Out-of-Sample Evaluation}
\label{algo:ew}
\end{algorithm}

This procedure ensures that the prediction target is never used for model selection or early stopping, preventing data leakage. The internal validation split (last 20\% of the training window) is used by the MLP for early stopping. For XGBoost and Elastic Net, which use internal cross-validation, the validation split is not used. The model architecture is selected on the initial training window and held fixed as the window expands; at each subsequent step, only the model weights are re-estimated on the expanded training data. This is consistent with standard practice in the forecasting literature, where model specification is not revised at each evaluation step.

\subsection{Monthly GDP Estimation and Reconciliation}

\begin{algorithm}[H]
\SetAlgoLined
\KwData{Trained model $\hat{f}(\cdot; \hat{\theta})$; monthly data \texttt{X\_prime\_m\_df}; quarterly GDP growth $Y_q$}
\KwResult{Reconciled monthly GDP growth estimates $\hat{y}_m$ and levels}

\tcp{Step 1: Generate preliminary monthly signal}
$\tilde{y}_m \leftarrow \hat{f}(X_m; \hat{\theta})$ for all months $m$\;

\tcp{Step 2: Construct MA(5) constraint matrix $M$}
\For{each quarter $q$ with last month $m$}{
    Row of $M$: weights $[\frac{1}{3}, \frac{2}{3}, 1, \frac{2}{3}, \frac{1}{3}]$ at positions $[m, m{-}1, m{-}2, m{-}3, m{-}4]$\;
}

\tcp{Step 3: Reconcile with quarterly constraints}
$z \leftarrow$ vector of observed quarterly growth rates\;
$\hat{y} \leftarrow \tilde{y} + M^\top(MM^\top)^{-1}(z - M\tilde{y})$ \tcp*{Minimum-norm adjustment}

\tcp{Step 4: Recover monthly GDP levels}
\For{$m = 1$ to $T$}{
    $\text{Level}_m \leftarrow \text{Level}_{m-1} \times \exp(\hat{y}_m)$\;
}

\tcp{Step 5: Compute reconciliation diagnostics}
\For{each quarter $q$}{
    $k_q \leftarrow Y_q \; / \; (M \tilde{y})_q$ \tcp*{Adjustment factor}
}

\caption{Monthly GDP Estimation with Mariano-Murasawa Reconciliation}
\label{algo:reconciliation}
\end{algorithm}

The reconciliation in Step~3 replaces the simpler proportional \citet{denton1971adjustment} method used in some applied work. The Mariano-Murasawa constraint reflects the log-linear approximation relating monthly and quarterly growth rates via a five-term moving average (see \Cref{app:reconciliation} for the full derivation). The adjustment factor $k_q$ in Step~5 serves as a diagnostic: values close to unity indicate that the model's raw monthly signal is already well-calibrated.

\newpage
\section{Reconciliation Details \label{app:reconciliation}}

\renewcommand{\thetable}{E\arabic{table}}\setcounter{table}{0}
\renewcommand{\thefigure}{E\arabic{figure}}\setcounter{figure}{0}
\renewcommand\theequation{E\arabic{equation}}\setcounter{equation}{0}

This section derives the reconciliation procedure described in \Cref{sub:framework}. The derivation follows \citet{mariano2003new, mariano2010coincident}.

\subsection{Temporal Aggregation Constraint}

Let $y_m$ denote the unobserved monthly log-difference of GDP for month $m$, and let $Y_q$ denote the observed quarterly log-difference. Under the log-linear approximation, the relationship between monthly and quarterly growth rates is:
\begin{flalign}
    Y_q = \frac{1}{3} y_m + \frac{2}{3} y_{m-1} + y_{m-2} + \frac{2}{3} y_{m-3} + \frac{1}{3} y_{m-4},
\end{flalign}
where $m$ denotes the last month of quarter $q$. This five-term moving average arises from the approximation $\log(\text{GDP}_q) \approx \log\!\left(\frac{1}{3}\sum_{j \in q} \text{GDP}_{m_j}\right)$, which relates the quarterly level to the geometric mean of monthly levels.

Stacking these constraints across all $Q$ quarters yields the matrix equation:
\begin{flalign}
    M y = z,
\end{flalign}
where $y \in \mathbb{R}^T$ is the vector of monthly growth rates, $z \in \mathbb{R}^Q$ is the vector of observed quarterly growth rates, and $M \in \mathbb{R}^{Q \times T}$ is the constraint matrix. Each row of $M$ contains the weights $[\frac{1}{3}, \frac{2}{3}, 1, \frac{2}{3}, \frac{1}{3}]$ placed at the appropriate monthly positions, with zeros elsewhere.

\subsection{Minimum-Norm Reconciliation}

Given a preliminary monthly signal $\tilde{y} \in \mathbb{R}^T$ from the machine learning model, we seek adjusted estimates $\hat{y}$ that satisfy the quarterly constraints exactly while deviating minimally from the preliminary signal. This is formulated as the constrained optimization problem:
\begin{flalign}
    \min_{y \in \mathbb{R}^T} \quad & \| y - \tilde{y} \|^2 \\
    \text{s.t.} \quad & M y = z.
\end{flalign}

The Lagrangian is:
\begin{flalign}
    \mathcal{L}(y, \lambda) = (y - \tilde{y})^\top (y - \tilde{y}) + \lambda^\top (z - My),
\end{flalign}
where $\lambda \in \mathbb{R}^Q$ is the vector of Lagrange multipliers. The first-order conditions are:
\begin{flalign}
    \frac{\partial \mathcal{L}}{\partial y} &= 2(y - \tilde{y}) - M^\top \lambda = 0, \\
    \frac{\partial \mathcal{L}}{\partial \lambda} &= z - My = 0.
\end{flalign}

From the first condition:
\begin{flalign}
    y = \tilde{y} + \tfrac{1}{2} M^\top \lambda.
\end{flalign}

Substituting into the constraint:
\begin{flalign}
    M\!\left(\tilde{y} + \tfrac{1}{2} M^\top \lambda\right) = z \quad \Longrightarrow \quad \lambda = 2(MM^\top)^{-1}(z - M\tilde{y}).
\end{flalign}

The closed-form solution for the reconciled monthly estimates is therefore:
\begin{flalign}
    \hat{y} = \tilde{y} + M^\top (MM^\top)^{-1} (z - M\tilde{y}),
\end{flalign}
which is the minimum-norm adjustment that satisfies $M\hat{y} = z$ exactly. This can be verified by substituting back:
\begin{flalign}
    M\hat{y} &= M\tilde{y} + MM^\top(MM^\top)^{-1}(z - M\tilde{y}) = M\tilde{y} + z - M\tilde{y} = z. \quad \checkmark
\end{flalign}

\subsection{Comparison with Proportional Denton Method}

The proportional \citet{denton1971adjustment} method assumes the simpler aggregation constraint $\sum_{m \in q} y_m = Y_q$, implemented via:
\begin{flalign}
    J = I_Q \otimes \mathbf{1}_3^\top,
\end{flalign}
where $I_Q$ is the $Q \times Q$ identity matrix, $\mathbf{1}_3^\top = (1, 1, 1)$, and $\otimes$ denotes the Kronecker product. Under proportional Denton, the adjusted estimates are $\hat{y}_m = k_q \cdot \tilde{y}_m$ where $k_q = Y_q / \sum_{j \in q} \tilde{y}_j$.

The Mariano-Murasawa MA(5) formulation used in this paper replaces $J$ with $M$, which encodes the weights $[\frac{1}{3}, \frac{2}{3}, 1, \frac{2}{3}, \frac{1}{3}]$ and allows for cross-quarter dependencies. This is the standard approximation in the mixed-frequency literature and is consistent with the treatment in \citet{koop2023reconciled} and \citet{chan2023high}.

\newpage
\section{Additional Results \label{app:additional}}
\renewcommand{\thetable}{F\arabic{table}} \setcounter{table}{0}
\renewcommand{\thefigure}{F\arabic{figure}} \setcounter{figure}{0}
\renewcommand\theequation{F\arabic{equation}} \setcounter{equation}{0}

This appendix reports results for all four models (Chow-Lin, Elastic Net, XGBoost, MLP) across all countries and lag specifications.

\Cref{tab:full_results} presents the complete expanding-window out-of-sample metrics for all 48 model-country-lag configurations. \Cref{tab:full_dm} reports all 72 pairwise Diebold-Mariano tests. \Cref{tab:lag_sensitivity} summarizes the best model at each lag specification for each country.

\begin{longtable}{ll c ccccc}
\caption{Expanding-Window Out-of-Sample Results: All Countries, Models, and Lag Specifications}
\label{tab:full_results} \\
\toprule
Country & Model & Lag & RMSE & MAE & $R^2$ & Corr. & Sign Acc. \\
\midrule
\endfirsthead
\toprule
Country & Model & Lag & RMSE & MAE & $R^2$ & Corr. & Sign Acc. \\
\midrule
\endhead
United States & Chow-Lin & 0 & \textbf{0.0085} & 0.0053 & 0.723 & 0.866 & 92.5\% \\
 & Elastic Net & 0 & 0.0115 & 0.0061 & 0.488 & 0.785 & 96.2\% \\
 & MLP & 0 & 0.0123 & 0.0069 & 0.414 & 0.657 & 96.2\% \\
 & XGBoost & 0 & 0.0153 & 0.0063 & 0.103 & 0.370 & 94.3\% \\
 & Elastic Net & 1 & \textbf{0.0058} & 0.0040 & 0.870 & 0.942 & 96.2\% \\
 & Chow-Lin & 1 & 0.0138 & 0.0064 & 0.271 & 0.527 & 83.0\% \\
 & XGBoost & 1 & 0.0155 & 0.0065 & 0.074 & 0.298 & 94.3\% \\
 & MLP & 1 & 0.0255 & 0.0167 & $-$1.496 & 0.487 & 75.5\% \\
 & Elastic Net & 2 & \textbf{0.0078} & 0.0050 & 0.771 & 0.897 & 96.2\% \\
 & XGBoost & 2 & 0.0151 & 0.0062 & 0.138 & 0.431 & 94.2\% \\
 & MLP & 2 & 0.0169 & 0.0074 & $-$0.085 & -0.058 & 88.5\% \\
 & Chow-Lin & 2 & 0.0234 & 0.0098 & $-$1.069 & 0.365 & 86.5\% \\
\addlinespace
Germany & Elastic Net & 0 & \textbf{0.0160} & 0.0068 & 0.225 & 0.521 & 74.1\% \\
 & MLP & 0 & 0.0165 & 0.0075 & 0.168 & 0.410 & 64.8\% \\
 & Chow-Lin & 0 & 0.0165 & 0.0076 & 0.167 & 0.410 & 77.8\% \\
 & XGBoost & 0 & 0.0170 & 0.0076 & 0.124 & 0.375 & 72.2\% \\
 & MLP & 1 & \textbf{0.0150} & 0.0078 & 0.313 & 0.567 & 74.1\% \\
 & Chow-Lin & 1 & 0.0160 & 0.0074 & 0.220 & 0.499 & 66.7\% \\
 & Elastic Net & 1 & 0.0163 & 0.0072 & 0.190 & 0.612 & 70.4\% \\
 & XGBoost & 1 & 0.0170 & 0.0072 & 0.121 & 0.367 & 74.1\% \\
 & Chow-Lin & 2 & \textbf{0.0165} & 0.0081 & 0.171 & 0.417 & 70.4\% \\
 & XGBoost & 2 & 0.0166 & 0.0071 & 0.166 & 0.457 & 68.5\% \\
 & Elastic Net & 2 & 0.0182 & 0.0080 & $-$0.005 & -0.044 & 68.5\% \\
 & MLP & 2 & 0.0191 & 0.0095 & $-$0.110 & 0.116 & 70.4\% \\
\addlinespace
United Kingdom & MLP & 0 & \textbf{0.0368} & 0.0135 & 0.092 & 0.314 & 83.3\% \\
 & Chow-Lin & 0 & 0.0375 & 0.0115 & 0.058 & 0.272 & 74.1\% \\
 & Elastic Net & 0 & 0.0378 & 0.0119 & 0.042 & 0.232 & 79.6\% \\
 & XGBoost & 0 & 0.0381 & 0.0120 & 0.029 & 0.187 & 83.3\% \\
 & Chow-Lin & 1 & \textbf{0.0372} & 0.0116 & 0.074 & 0.331 & 83.3\% \\
 & XGBoost & 1 & 0.0378 & 0.0123 & 0.045 & 0.246 & 85.2\% \\
 & Elastic Net & 1 & 0.0380 & 0.0118 & 0.030 & 0.201 & 85.2\% \\
 & MLP & 1 & 0.0382 & 0.0170 & 0.023 & 0.246 & 75.9\% \\
 & Chow-Lin & 2 & \textbf{0.0374} & 0.0129 & 0.061 & 0.281 & 74.1\% \\
 & XGBoost & 2 & 0.0377 & 0.0112 & 0.049 & 0.312 & 87.0\% \\
 & Elastic Net & 2 & 0.0380 & 0.0116 & 0.032 & 0.221 & 87.0\% \\
 & MLP & 2 & 0.0384 & 0.0126 & 0.013 & 0.129 & 87.0\% \\
\addlinespace
China & Elastic Net & 0 & \textbf{0.1143} & 0.0939 & 0.078 & 0.457 & 79.2\% \\
 & Chow-Lin & 0 & 0.1143 & 0.0967 & 0.077 & 0.278 & 77.1\% \\
 & XGBoost & 0 & 0.1160 & 0.0930 & 0.049 & 0.287 & 75.0\% \\
 & MLP & 0 & 0.1300 & 0.1075 & $-$0.193 & 0.078 & 62.5\% \\
 & XGBoost & 1 & \textbf{0.1164} & 0.0938 & 0.043 & 0.246 & 72.9\% \\
 & MLP & 1 & 0.1168 & 0.0889 & 0.036 & 0.277 & 68.8\% \\
 & Chow-Lin & 1 & 0.1193 & 0.0978 & $-$0.005 & 0.337 & 62.5\% \\
 & Elastic Net & 1 & 0.1201 & 0.0912 & $-$0.019 & -0.034 & 75.0\% \\
 & XGBoost & 2 & \textbf{0.1172} & 0.0959 & 0.044 & 0.244 & 76.6\% \\
 & MLP & 2 & 0.1177 & 0.0924 & 0.036 & 0.458 & 80.9\% \\
 & Elastic Net & 2 & 0.1209 & 0.0930 & $-$0.018 & -0.064 & 74.5\% \\
 & Chow-Lin & 2 & 0.1611 & 0.1285 & $-$0.807 & 0.375 & 61.7\% \\
\addlinespace
\bottomrule
\end{longtable}

\begin{longtable}{ll ccc c}
\caption{Diebold-Mariano Test Results: All Pairwise Comparisons}
\label{tab:full_dm} \\
\toprule
Country & Lag & Model 1 & Model 2 & DM Stat & $p$-value \\
\midrule
\endfirsthead
\toprule
Country & Lag & Model 1 & Model 2 & DM Stat & $p$-value \\
\midrule
\endhead
US & 0 & Chow-Lin & XGBoost & -1.308 & 0.1910 \\
 & 0 & Chow-Lin & Elastic Net & -1.175 & 0.2402 \\
 & 0 & Chow-Lin & MLP & -1.110 & 0.2669 \\
 & 0 & Elastic Net & XGBoost & -1.089 & 0.2763 \\
 & 0 & XGBoost & MLP & +0.759 & 0.4479 \\
 & 0 & Elastic Net & MLP & -0.701 & 0.4835 \\
 & 1 & Chow-Lin & MLP & -3.498 & 0.0005$^{***}$ \\
 & 1 & Elastic Net & MLP & -3.155 & 0.0016$^{***}$ \\
 & 1 & XGBoost & MLP & -2.545 & 0.0109$^{**}$ \\
 & 1 & Elastic Net & XGBoost & -1.437 & 0.1508 \\
 & 1 & Chow-Lin & Elastic Net & +1.362 & 0.1733 \\
 & 1 & Chow-Lin & XGBoost & -0.578 & 0.5630 \\
 & 2 & XGBoost & MLP & -1.674 & 0.0941$^{*}$ \\
 & 2 & Elastic Net & MLP & -1.408 & 0.1592 \\
 & 2 & Chow-Lin & Elastic Net & +1.388 & 0.1652 \\
 & 2 & Elastic Net & XGBoost & -1.314 & 0.1890 \\
 & 2 & Chow-Lin & XGBoost & +1.053 & 0.2922 \\
 & 2 & Chow-Lin & MLP & +0.868 & 0.3856 \\
\addlinespace
Germany & 0 & Chow-Lin & Elastic Net & +1.798 & 0.0721$^{*}$ \\
 & 0 & Elastic Net & XGBoost & -1.411 & 0.1583 \\
 & 0 & Elastic Net & MLP & -1.120 & 0.2627 \\
 & 0 & Chow-Lin & XGBoost & -0.700 & 0.4840 \\
 & 0 & XGBoost & MLP & +0.683 & 0.4944 \\
 & 0 & Chow-Lin & MLP & +0.018 & 0.9858 \\
 & 1 & Elastic Net & XGBoost & -1.484 & 0.1379 \\
 & 1 & Chow-Lin & XGBoost & -1.295 & 0.1955 \\
 & 1 & Chow-Lin & Elastic Net & -0.861 & 0.3894 \\
 & 1 & XGBoost & MLP & +0.831 & 0.4057 \\
 & 1 & Elastic Net & MLP & +0.613 & 0.5399 \\
 & 1 & Chow-Lin & MLP & +0.506 & 0.6129 \\
 & 2 & XGBoost & MLP & -2.549 & 0.0108$^{**}$ \\
 & 2 & Chow-Lin & MLP & -1.946 & 0.0516$^{*}$ \\
 & 2 & Elastic Net & XGBoost & +1.134 & 0.2567 \\
 & 2 & Chow-Lin & Elastic Net & -0.967 & 0.3335 \\
 & 2 & Elastic Net & MLP & -0.651 & 0.5150 \\
 & 2 & Chow-Lin & XGBoost & -0.116 & 0.9080 \\
\addlinespace
UK & 0 & Chow-Lin & Elastic Net & -1.366 & 0.1719 \\
 & 0 & Chow-Lin & XGBoost & -1.085 & 0.2779 \\
 & 0 & Elastic Net & XGBoost & -0.393 & 0.6941 \\
 & 0 & XGBoost & MLP & +0.328 & 0.7427 \\
 & 0 & Elastic Net & MLP & +0.309 & 0.7575 \\
 & 0 & Chow-Lin & MLP & +0.200 & 0.8414 \\
 & 1 & Chow-Lin & Elastic Net & -0.973 & 0.3304 \\
 & 1 & Chow-Lin & XGBoost & -0.675 & 0.4998 \\
 & 1 & Elastic Net & XGBoost & +0.606 & 0.5444 \\
 & 1 & Chow-Lin & MLP & -0.483 & 0.6290 \\
 & 1 & XGBoost & MLP & -0.163 & 0.8707 \\
 & 1 & Elastic Net & MLP & -0.049 & 0.9611 \\
 & 2 & Chow-Lin & MLP & -1.668 & 0.0954$^{*}$ \\
 & 2 & XGBoost & MLP & -1.668 & 0.0954$^{*}$ \\
 & 2 & Elastic Net & XGBoost & +0.996 & 0.3195 \\
 & 2 & Chow-Lin & Elastic Net & -0.762 & 0.4459 \\
 & 2 & Elastic Net & MLP & -0.653 & 0.5140 \\
 & 2 & Chow-Lin & XGBoost & -0.551 & 0.5814 \\
\addlinespace
China & 0 & Elastic Net & MLP & -1.369 & 0.1709 \\
 & 0 & Chow-Lin & MLP & -1.343 & 0.1792 \\
 & 0 & XGBoost & MLP & -1.145 & 0.2522 \\
 & 0 & Elastic Net & XGBoost & -0.659 & 0.5100 \\
 & 0 & Chow-Lin & XGBoost & -0.364 & 0.7161 \\
 & 0 & Chow-Lin & Elastic Net & +0.015 & 0.9880 \\
 & 1 & Elastic Net & XGBoost & +1.304 & 0.1924 \\
 & 1 & Elastic Net & MLP & +0.395 & 0.6927 \\
 & 1 & Chow-Lin & XGBoost & +0.258 & 0.7966 \\
 & 1 & Chow-Lin & MLP & +0.219 & 0.8270 \\
 & 1 & Chow-Lin & Elastic Net & -0.063 & 0.9496 \\
 & 1 & XGBoost & MLP & -0.056 & 0.9554 \\
 & 2 & Chow-Lin & XGBoost & +2.364 & 0.0181$^{**}$ \\
 & 2 & Chow-Lin & MLP & +2.322 & 0.0202$^{**}$ \\
 & 2 & Chow-Lin & Elastic Net & +2.081 & 0.0374$^{**}$ \\
 & 2 & Elastic Net & XGBoost & +1.235 & 0.2167 \\
 & 2 & Elastic Net & MLP & +0.460 & 0.6454 \\
 & 2 & XGBoost & MLP & -0.075 & 0.9405 \\
\addlinespace
\bottomrule
\multicolumn{6}{p{0.85\textwidth}}{\footnotesize \textit{Notes:} Diebold-Mariano test with Newey-West HAC variance and squared error loss. A negative DM statistic indicates that Model~1 has lower average loss (Model~1 is more accurate); a positive statistic favors Model~2. $^{*}$, $^{**}$, $^{***}$ denote significance at the 10\%, 5\%, and 1\% levels, respectively.} \\
\end{longtable}

\begin{table}[htbp]
\centering
\caption{Lag Sensitivity: Best Model per Country and Lag}
\label{tab:lag_sensitivity}
\begin{tabular}{ll ccccc}
\toprule
Country & Lag & Best Model & RMSE & $R^2$ & Corr. & Sign Acc. \\
\midrule
United States & 0 & Chow-Lin & 0.0085 & 0.723 & 0.866 & 92.5\% \\
 & 1 & Elastic Net & 0.0058 & 0.870 & 0.942 & 96.2\% \\
 & 2 & Elastic Net & 0.0078 & 0.771 & 0.897 & 96.2\% \\
\addlinespace
Germany & 0 & Elastic Net & 0.0160 & 0.225 & 0.521 & 74.1\% \\
 & 1 & MLP & 0.0150 & 0.313 & 0.567 & 74.1\% \\
 & 2 & Chow-Lin & 0.0165 & 0.171 & 0.417 & 70.4\% \\
\addlinespace
United Kingdom & 0 & MLP & 0.0368 & 0.092 & 0.314 & 83.3\% \\
 & 1 & Chow-Lin & 0.0372 & 0.074 & 0.331 & 83.3\% \\
 & 2 & Chow-Lin & 0.0374 & 0.061 & 0.281 & 74.1\% \\
\addlinespace
China & 0 & Elastic Net & 0.1143 & 0.078 & 0.457 & 79.2\% \\
 & 1 & XGBoost & 0.1164 & 0.043 & 0.246 & 72.9\% \\
 & 2 & XGBoost & 0.1172 & 0.044 & 0.244 & 76.6\% \\
\addlinespace
\bottomrule
\end{tabular}
\end{table}

\end{document}